%% file: main.tex
\documentclass[]{elsarticle}

\usepackage{hyperref}

\journal{CMAME}
\usepackage[american]{babel}
\usepackage[utf8]{inputenc}
\usepackage[T1]{fontenc}

\usepackage{amsmath}
\usepackage{amssymb}
\usepackage{booktabs}
\usepackage[dvipsnames]{xcolor}
\usepackage{comment}
\usepackage{graphicx}
\usepackage{mathtools}
\usepackage{subfig}

\usepackage{tikz,tikz-3dplot}
\usetikzlibrary
   {arrows,calc,through,backgrounds,matrix,positioning,decorations.pathmorphing,shapes.geometric,3d}

\usepackage{gnuplot-lua-tikz}

\newcommand{\e}{\varepsilon}

\newcommand{\er}{\varepsilon_{r}\,}

\newcommand{\mur}{\mu_{r}\,}
\newcommand{\muri}{\mu_{r}^{-1}\,}

\renewcommand{\vec}[1]{\boldsymbol{#1}}

\newcommand{\p}{\partial}

\newcommand{\J}{\mathcal{J}}
\newcommand{\ssr}{\sigma_r}
\newcommand{\vB}{\vec B}
\newcommand{\vE}{\vec E}
\newcommand{\vH}{\vec H}
\newcommand{\vJa}{\vec J_{a}}

\newcommand{\vMF}{\vec{\mathcal{F}}}
\newcommand{\vn}{\vec \nu}
\newcommand{\vp}{\vec \varphi}
\newcommand{\vx}{\vec x}
\newcommand{\vZ}{\vec Z}

\newcommand{\RN}[1]{%
  \textup{\uppercase\expandafter{\romannumeral#1}}%
}

\newcommand{\sgn}{\text{sgn}}
\newcommand{\TTH}{\mathbb{T}_{\mskip-\medmuskip H}}

\newcommand{\dx}{\,{\mathrm d}x}
\newcommand{\dy}{\,{\mathrm d}y}
\newcommand{\dox}{\,{\mathrm d}o_x}

\begin{document}

\begin{frontmatter}

\title{%
  Adaptive finite element simulations of waveguide configurations %
  involving parallel 2D material sheets}

\author[minnesota]{Jung Heon Song}
\ead{songx762@umn.edu}

\author[texas]{Matthias Maier\corref{corresponding}}
\ead{maier@math.tamu.edu}

\author[minnesota]{Mitchell Luskin}
\ead{luskin@umn.edu}

\address[minnesota]{%
  School of Mathematics, University of Minnesota, %
  206 Church Street SE, Minneapolis, MN 55455, USA.}

\address[texas]{%
  Department of Mathematics, Texas A\&M University, %
  3368 TAMU, College Station, TX 77843, USA.}

\cortext[corresponding]{Corresponding author}

\begin{abstract}
  We discuss analytically and numerically the propagation and energy
  transmission of electromagnetic waves caused by the coupling of surface
  plasmon polaritons (SPPs) between two spatially separated layers of 2D
  materials, such as graphene, at subwavelength distances. We construct an
  adaptive finite-element method to compute the ratio of energy transmitted
  within these waveguide structures reliably and efficiently. At its heart,
  the method is built upon a goal-oriented a posteriori error estimation
  with the dual-weighted residual method (DWR).

  Furthermore, we derive analytic solutions of the two-layer system,
  compare those to (known) single-layer configurations, and compare and
  validate our numerical findings by comparing numerical and analytical
  values for optimal spacing of the two-layer configuration. Additional
  aspects of our numerical treatment, such as local grid refinement, and
  the utilization of perfectly matched layers (PMLs) are examined in
  detail.
\end{abstract}


\end{frontmatter}


\section{Introduction}

Graphene is a two-dimensional carbon allotrope with one-atomic thickness
that is arranged in a honeycomb lattice structure~\cite{geim04}. It has a
wide potential for applications in nanophotonics due to a number of
desirable electronic and optical features, such as extreme confinement, low
losses, and tunability~\cite{geim04,bludov13}. In the infrared regime, the
\emph{electric surface conductivity} of such a 2D material is characterized
by being complex-valued with a dominant positive imaginary part. This
allows for the propagation of \emph{surface plasmon polaritons} (SPPs),
which are highly confined to the 2D material and slowly decaying
electromagnetic waves. Within the frequency domain of interest, the
wavelength of a SPP is up to two orders of magnitude smaller than that of
the exciting ambient light.

Waveguide structures that enable subwavelength confinement of the
optical modes are of great importance in
nanophotonics~\cite{samaier07,gramotnev10,brongersma00,oulton08}.
Traditionally, waveguides have been implemented through local modulation of
the shape and/or refractive index profile of the optical dielectric
medium~\cite{collins-book}. These dielectric waveguides, however, are
restricted by the diffraction limit of light, $\lambda_0 / n$, where
$\lambda_0$ is the wavelength in free space and $n$ is the refractive
index~\cite{gramotnev10}. SPPs, on the other hand, can be confined
within a very small area beyond the diffraction limit of light, and can be
used as an information carrier for highly-integrated photonic
circuits~\cite{bde03}. Thus, a number of plasmon-based waveguides have been
proposed in the past decade, such as metallic nanowires~\cite{weeber99,
krenn02}, metallic nanoparticle arrays~\cite{brongersma00, maier03a},
hybrid plasmonic waveguides~\cite{oulton08}, and gain assisted plasmonic
waveguides~\cite{nezhad04}.

Although some studies that investigated the plasma modes and optical SPP
modes of a \emph{double-layer} graphene were
published~\cite{hwang09,hwang10, stauber11}, the majority of research on
graphene has been on a single-layer systems consisting of a single, planar
sheet of graphene~\cite{neto09,fei11,ju11}. By introducing a second,
parallel sheet, placed at a small but finite distance to the other sheet,
it is possible to drastically change the confinement and propagation
characteristics of SPPs. The purpose of this paper is two-fold.

First, we numerically investigate an infinite 2D waveguide by computing a
finite element approximation for the solution of the corresponding
scattering problem governed by time-harmonic Maxwell's equations.
This prototypical geometry is motivated by proposed waveguide
configurations that include graphene layers, or carbon nanotubes as an
integral part of their design~\cite{prb2009}.
By adjusting the confinement of the two-layer system, we find an optimal
spacing for which the coupling of the SPPs is maximal. To this end, a
goal-oriented mesh refinement strategy and a perfectly matched layer are
utilized.

Secondly, we derive and discuss an integral equation describing the
time-harmonic electromagnetic field of a double-layer system. We examine
the contributions from a pole of the scattered field solution, which are
responsible for the generation of SPPs. We demonstrate that our findings
are in accordance with those from a single-layer system by observing the
evolution of the scattered field under different interlayer spacings.
In conclusion, we note that the maximal spacing found in the numerics is in
agreement with the value found in the analytical expressions.


\subsection{Related works}

The SPP dispersion of a single graphene layer and a single graphene layer
deposited on dielectric substrates has been extensively investigated by
many authors~\cite{wang2012,neto09,ju11}. Additional confinement and a
change of propagation characteristics can be achieved by stacking another
layer of these 2D materials on top of a single-layer~\cite{gan12}. By
extension, theoretical aspects of bilayer
graphene~\cite{stauber11,hwang07,hwang09}, multilayer
graphene~\cite{falkovsky07}, and intercalated graphite~\cite{shung86} have
been studied recently. For example,~\cite{stauber11} finds a frequency for
which reflection in a double-layer graphene system, with both equal and
different surface conductivities, is zero, leading to exponentially
amplified transmitted modes.

\cite{maier18a,maierxxa} shows that for plasmonic crystals, which consist
of stacked metallic layers arranged periodically with subwavelength
distance, embedded in a dielectric medium, the TM polarized waves
experience an effective dielectric function that combines a bulk energy of
the microstructure of the ambient dielectric medium and surface average of
the surface conductivity of each sheet. Homogenization of layered
structures and extension to a general hypersurface are also discussed.

However, a rigorous numerical and analytical treatment of waveguide
configurations involving time-harmonic Maxwell's equations is not of
primary interests in these publications.
This paper aims to address three points. First, we introduce a
reliable and efficient numerical method to readily compute propagation
characteristics of the SPPs in a double-layer structure. Second, we
validate the numerical findings against an analytical solution. Third, we
demonstrate that our numerical approach is easily extensible to different
computational domains.


\subsection{Paper organization}

The paper is organized as follows. In Section~\ref{sec:variational}, we
derive the variational formulation that serves as the basis for our
numerical and analytical investigation. In Section~\ref{sec:numerics} we
develop the numerical framework, including a goal-oriented mesh adaptation
based on the dual weighted residual (DWR) method, and a perfectly matched
layer (PML). We perform a direct numerical simulation of a prototypical
two-layer system in Section~\ref{sec:dns}, and identify optimal spacings of
two-layer systems for maximal transmission. In
Section~\ref{sec:analytical}, we derive analytic solutions for the
two-layer system and validate our numerical findings against them. Finally,
Section~\ref{sec:conclusion} concludes the paper with a summary of our
results and an outlook.


\section{Variational formulation}
\label{sec:variational}

In this section, we lay out the variational formulation for
time-harmonic Maxwell's equations with an interface condition.
We introduce a rescaling for time-harmonic Maxwell's equations that
will ease the numerical computation of SPPs~\cite{maier17a}.


\subsection{Preliminaries: Maxwell's equations}

Time-harmonic Maxwell's equations with an electric current
density read ~\cite{monk03, stratton41}:
\begin{align}
  \begin{cases}
    \begin{aligned}
      -i\omega\,\vB(\vx)+\nabla\times\vE(\vx) \;&=\; 0,
      \\[0.3em]
      \nabla\cdot \vB(\vx) \;&=\; 0,
      \\[0.3em]
      i\omega\e(\vx)\vE(\vx)+\nabla\times\big(\mu(\vx)^{-1}\vB(\vx)\big) \;&=\;
      \vJa(\vx),
      \\[0.3em]
      \nabla\cdot\big(\e(\vx)\vE(\vx)\big) \;&=\; \frac 1{i\omega}
      \nabla\cdot\vJa(\vx).
    \end{aligned}
  \end{cases}
  \label{eq:timeharmonicmaxwell}
\end{align}
Here, $\vE(\vx)$ and $\vB(\vx)$ denote the electric and magnetic field,
respectively. $\mu(\vx)$ and $\e(\vx)$ are complex-valued rank 2 tensor
quantities, where $\mu(\vx)$ denotes the magnetic permeability and
$\e(\vx)$ denotes the electric permittivity. The vector-valued quantity
$\vJa(\vx)$ is an externally applied, electric current density. We assume a
general $\vx$ dependence of all quantities with some (weak) regularity
conditions to ensure unique solvability that will be stated later. The
(constant) \emph{temporal frequency} $\omega > 0$ arises from the
time-harmonic nature of the solution fields, i.\,e., a solution of
(\ref{eq:timeharmonicmaxwell}) is a special solution of the general
\emph{time-dependent Maxwell's equations} by rewriting all vector-valued
components $\vMF$ by
\begin{align}
  \vMF(\vx,t) = \text{Re}\,\big(e^{-i\omega t}\vMF(\vx)\big).
\end{align}

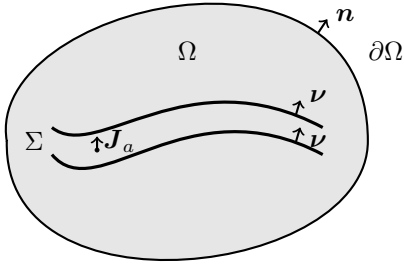
\begin{figure}[t]
  \centering
  \vspace{-1em}
  \begin{tikzpicture}[
    scale=1.2,
    extended/.style={shorten >=-#1, shorten <=-#1},
    extended/.default=1cm]
    \node at (-0.0, 1.0) {$\Omega$};
    \path [fill=black, opacity=0.1]
      (-2, 0) .. controls (-2.05, -2.0) and (2, -1.5) ..
      ( 2, 0) .. controls ( 2,  2.5) and (-2.25, 1.5) .. cycle;
    \path [thick, draw]
      (-2, 0) .. controls (-2.05, -2.0) and (2, -1.5) ..
      ( 2, 0) .. controls ( 2,  2.5) and (-2.25, 1.5) .. cycle;
    \path [very thick,draw]
      (-1.5, 0.15) .. controls (-1, -0.25) and (0.00, 0.95) .. (1.5, 0.15);
    \path [very thick,draw]
      (-1.5, -0.15) .. controls (-1, -0.75) and (0.00, 0.65) .. (1.5, -0.15);
    \node at (-1.70, 0.0) {$\Sigma$};
    \path [thick, ->, draw] (1.2, 0.28) -- (1.25, 0.45);
    \node at (1.45, 0.50) {$\vn$};
    \path [thick, ->, draw] (1.2, -0.02) -- (1.25, 0.15);
    \node at (1.45, 0.0) {$\vn$};
    \path [thick, ->, draw] (-1.0, -0.1) -- (-1.0, 0.05);
    \node at (-0.75, 0.0) {$\vJa$};
    \draw[fill] (-1.0, -0.1) circle (0.6pt);
    \path [thick, ->, draw] (1.45, 1.2) -- (1.55, 1.35);
    \node at (1.75, 1.4) {$\vec n$};
    \node at (2.2, 1.0) {$\partial\Omega$};
  \end{tikzpicture}
  \vspace{-2em}
  \caption{
    Schematic of the computational domain, $\Omega$, with boundary
    $\partial\Omega$, outer normal $\vec n$, and normal field $\vec\nu$
    defined on the waveguide $\Sigma$. An electric Hertzian dipole, $\vJa$,
    is placed inside a prescribed waveguide structure, $\Sigma$.}
  \label{fig:domain}
\end{figure}

We are interested in simulating waveguide configurations that feature 2D
material sheets. The sheets are modeled as an idealized hypersurface
$\Sigma$ with an effective surface conductivity $\sigma(\vx)$ defined on
$\Sigma$~\cite{margetis15,hanson08,  bludov13}. In general, $\Sigma$  shall
consist of two parallel, possibly curved, conducting sheets separated by a
fixed distance $d$; see Figure~\ref{fig:domain}. The discontinuity along
the surface due to the conductivity leads to a jump condition in the
tangential component of the magnetic field
\cite{margetis15,monk03,bludov13}:
\begin{align}
  \begin{cases}
    \begin{aligned}
      \big[\vn\times(\mu^{-1}\vB)\big]_\Sigma \;&=\;
      \sigma(\vx)\vE_T\Big|_\Sigma,
      \\[0.3em]
      \big[\vn\times\vE\big]_\Sigma \;&=\; 0,
    \end{aligned}
  \end{cases}
  \label{eq:jumpcondition}
\end{align}
where $\vn$ is a fixed normal vector field associated with $\Sigma$; the
symbol $[\,.\,]_\Sigma$ denotes the jump over $\Sigma$ with respect to
$\vn$,
\begin{align}
  \big[\vMF\big]_\Sigma(x)
  :=\lim_{s\searrow0}\big(\vMF(x+s\vn)-\vMF(x-s\vn)\big),
\end{align}
and the subscript $T$ denotes the tangential part of the respective vector,
$\vMF_T = (\vn\times\vMF)\times\vn$. Under appropriate conditions on
$\sigma(\vx)$, jump condition \eqref{eq:jumpcondition} generates SPPs on
the interface~\cite{hanson08,bludov13}.

We make the assumption that $\e(\vx)$ and $\mu(\vx)$ become homogeneous and
isotropic for large $|\vx|$ and impose the Silver-M\"uller radiation
condition at infinity~\cite{maier17a}, viz.,
\begin{align}
  \lim_{|\vx|\to\infty} \{\vB\times\vx-c^{-1}|\vx|\vE\} = 0, \quad
  \lim_{|\vx|\to\infty} \{\vE\times\vx+c|\vx|\vB\} = 0, \quad
  \vx\not\in\Sigma.
  \label{eq:silver-muller}
\end{align}
Here, $c = 1 /\sqrt{\e\mu}$ denotes the speed of light. The explicit
inclusion of this condition is omitted in our numerical simulation by
incorporating an appropriate boundary condition and a PML.


\subsection{Rescaling and variational formulation}

Numerical values in SI units for solutions of
\eqref{eq:timeharmonicmaxwell} are many orders of magnitude apart. Further,
the typical length scale of SPP is one to two orders of magnitude smaller
than the free-space wavelength $k_0$~\cite{bludov13}. These discrepancies
in the magnitude of the length scales makes the direct numerical simulation
of the SPPs difficult. As a remedy we use a rescaling to dimensionless
units that normalizes the length scale by the free-space wavenumber, $k_0$,
\cite{maier17a}:
\begin{gather*}
  \vx \;\rightarrow\; k_0\,\vx,
  \quad
  \nabla \;\rightarrow\; \frac1{k_0}\,\nabla,
  \\
  \mu \;\rightarrow\; \mur =\frac1{\mu_0}\mu,
  \quad
  \e \;\rightarrow\; \er = \frac1{\e_0}\e,
  \quad
  \sigma \;\rightarrow\; \ssr =
  \sqrt{\frac{\mu_0}{\e_0}}\,\sigma.
\end{gather*}
Here, $\e_0$ and $\mu_0$ denote the vacuum permittivity and permeability,
respectively. This leads to two distinctly separated scales: one related to
the free-space wavenumber $\sim 1$, and another for the SPP wavenumber,
$k_{\text{SPP}}\sim1/\ssr$, on the conducting sheets~\cite{maier17a}. The
rescaled, dimensionless form of time-harmonic Maxwell's equation
\eqref{eq:timeharmonicmaxwell} reads
\begin{align}
  \nabla\times\big(\muri\nabla\times\vE\big)
  -\er\vE
  \;=\; i\,\vJa,
  \label{eq:2ndorderequationrescaled}
\end{align}
with the jump condition
\begin{align}
    \begin{aligned}
      \big[\vn\times(\muri\vB)]_\Sigma \;=\; \sigma_r\vE_T,
      \quad
      \big[\vn\times\vE]_\Sigma \;=\; 0.
    \end{aligned}
  \label{eq:jumpconditionrescaled}
\end{align}

Let the computational domain $\Omega\subset\mathbb{R}^3$ be bounded,
simply-connected, and Lipschitz-continuous with piecewise smooth boundary
$\p\Omega$. Assume that $\Sigma$ is a Lipschitz-continuous,
piecewise smooth boundary. The Silver-M\"uller radiation condition
(\ref{eq:silver-muller}) reads
\begin{align}
  \vn\times\vB + \sqrt{\muri\er}\vE_T = 0 \qquad (\vx\in\p\Omega).
  \label{eq:absorbingrescaled}
\end{align}
Multiplying (\ref{eq:2ndorderequationrescaled}) with the complex conjugate of
a test function $\vp$ and subsequent integration by parts using
(\ref{eq:jumpconditionrescaled}) and (\ref{eq:absorbingrescaled})
recovers the corresponding weak formulation:
\begin{align}
  \label{eq:variationalformulation}
  A(\vE,\vp)\;\;=\;\;i\int_\Omega\vJa\cdot\bar\vp\dx,
\end{align}
for $\vp\in X(\Omega) = \{\vp\in\vH(\text{curl};\Omega):\vp_T|_\Sigma\in
L^2(\Sigma)^3, \vp_T|_{\partial\Omega}\in L^2(\partial\Omega)^3\}$ and with
the bilinear form
\begin{multline}
  \label{eq:bilinear_form}
  A(\vec\psi,\vp) =
  \int_\Omega (\muri\nabla\times\vec\psi)\cdot(\nabla\times\bar\vp)
  - (\er\vec\psi)\cdot\bar\vp\,\dx
  \\
  - i\int_\Sigma(\ssr\vec\psi)\cdot\bar\vp\dox\; -
  i\int_{\p\Omega}\sqrt{\muri\er}(\vec\psi)\cdot\bar\vp\dox.
\end{multline}
In the above, $L^2(\cdot)^3$ denotes the space of vector-valued square
integrable functions, and $\vH(\text{curl})$ is the subspace of
$L^2(\cdot)^3$ consisting of square integrable functions whose
(distributive) curl admits a representation by a square integrable
function. Equation (\ref{eq:variationalformulation}) will serve as a
starting point for a finite element discretization. For elementary results
on existence and uniqueness, we refer
to~\cite{maier17a,monk03,colton83,colton98}.


\section{Numerics: Computational domain and discretization scheme}
\label{sec:numerics}

In this section, we briefly present the geometry and numerical tools
used in the computations. A PML is introduced and its role in
negating the undesired effects of the absorbing boundary condition of the
plasmon modes is described. Additionally, a local adaptive mesh refinement
strategy is presented that captures the highly oscillatory behaviors
of the plasmons near the interfaces. The variational
formulation~\eqref{eq:variationalformulation} is discretized on a
non-uniform quadrilateral mesh with higher-order, curl-conforming
N\'ed\'elec elements~\cite{nedelec01}. Such a choice is ideal where the
weak jump condition is naturally treated by aligning with the mesh. Let
$\vec X_h(\Omega) \subset \vec X(\Omega)$ be a finite element subspace
spanned by N\'ed\'elec elements. Then under a sufficiently refined initial
mesh, the variational formulation
\begin{align*}
  A(\vE_h,\vp)\;\;=\;\;i\int_\Omega\vJa\cdot\bar\vp\dx.
\end{align*}
is uniquely solvable for $\vE_h\in\vec X_h(\Omega)$ and for all
$\vp\in\vec X_h(\Omega)$.

\subsection{Geometry}

\begin{figure}[t]
    \subfloat[The computational domain. The location of the Hertzian dipole(s) is
    shown in the right.]{
    \begin{tikzpicture}[
      scale=1.2,
      extended/.style={shorten >=-#1, shorten <=-#1},
      extended/.default=1cm]
      \node at (-2.1, 2.1) {\small $\Omega$:};
      \path [fill=black, opacity=0.1] (-2.0,-2.0) -- (-2.0, 2.0) -- (2.0,2.0) -- (2.0,-2.0) -- cycle;
      \path [fill=white] (0,0) circle (1.5);
      \path [extended=0.5cm, thick, ->, opacity=0.75, draw] (-2, 0) -- (2, 0);
      \path [extended=0.5cm, thick, ->, opacity=0.75, draw] (0, -2) -- (0, 2);
      \node at (2.1, 0.1) {\small $x$};
      \node at (0.1, 2.1) {\small $y$};
      \path [thick, <->, draw] (-1.9,0.22) -- (-1.9, -0.22);
      \node at (-1.7, 0.05) {$d$};
      \path [-, dashed, draw] (-0.9, 2.0) -- (-.9, -2.0);
      \node at (-0.65, -.40) {$x_{\text{tr}}$};
      \path [-, dashed, draw] (+0.9, 2.0) -- (+.9, -2.0);
      \node at (+0.65, -.40) {$x_{\text{re}}$};
      \path [thick, draw, opacity=0.75] (0, -0.10) -- (0, 0.10);
      \path [thick, draw] (2, -0.1) -- (2, 0.1);
      \node at (2.20, -0.15) {\small $R$};
      \node at (1.40, -0.1) {\small $\rho$};
      \path [line width=0.5mm, draw] (-2.0, -0.22) -- (2.0, -0.22);
      \path [line width=0.5mm, draw] (-2.0, +0.22) -- (2.0, +0.22);
      \node at (1.2, +0.35) {\small $\Sigma_{12}$};
      \node at (1.2, -0.40) {\small $\Sigma_{23}$};
      \path [thick, opacity=0.75, dashed, draw] (0,0) circle (1.5);
      \node[rotate=-45.0] at (1.23,1.23) {\small PML};
      \node at (0,-2.5) {};
    \end{tikzpicture}
  }
  \begin{tabular}[b]{cc}
    \begin{tabular}[b]{c}
    \subfloat[Single dipole excitation.]{
    \begin{tikzpicture}[
      scale=1.2,
      extended/.style={shorten >=-#1, shorten <=-#1},
      extended/.default=1cm]
      \path [line width=0.5mm, draw] (-2.0, -0.42) -- (2.0, -0.42);
      \path [line width=0.5mm, draw] (-2.0, +0.42) -- (2.0, +0.42);
      \node at (1.2, +0.58) {\small $\Sigma_{12}$};
      \node at (1.2, -0.62) {\small $\Sigma_{23}$};
      \path [thick, <->, draw] (-1.9,0.42) -- (-1.9, -0.42);
      \node at (-1.7, 0.05) {$d$};
      \path [very thick, ->, draw] (-1.2, 0.0) -- (-1.2, 0.3);
      \draw [fill] (-1.2,0.0) circle (1.2pt);
      \node at (-0.9,0.0) {$\vJa$};
      \path [-, dashed, draw] (0.5, 1.0) -- (.5, -1.0);
      \node at (0.8, -.96) {$x_{\text{tr}}$};
    \end{tikzpicture}
    }
    \\
    \subfloat[Double dipole excitation.]{
    \begin{tikzpicture}[
      scale=1.2,
      extended/.style={shorten >=-#1, shorten <=-#1},
      extended/.default=1cm]
      \path [line width=0.5mm, draw] (-2.0, -0.42) -- (2.0, -0.42);
      \path [line width=0.5mm, draw] (-2.0, +0.42) -- (2.0, +0.42);
      \node at (1.2, +0.58) {\small $\Sigma_{12}$};
      \node at (1.2, -0.62) {\small $\Sigma_{23}$};
      \path [thick, <->, draw] (-1.9,0.42) -- (-1.9, -0.42);
      \node at (-1.7, 0.05) {$d$};
      \path [very thick, <-, draw] (-1.2,0.46) -- (-1.2,0.65);
      \path [very thick, <-, draw] (-1.2,0.38) -- (-1.2,0.17);
      \node at (-0.7, 0.58) {$\vJa$};
      \path [very thick, <-, draw] (-1.2,-0.65) -- (-1.2,-0.46);
      \path [very thick, <-, draw] (-1.2,-0.17) -- (-1.2,-0.38);
      \node at (-0.7, -0.26) {$\vJa$};
      \path [-, dashed, draw] (0.5, 1.0) -- (.5, -1.0);
      \node at (0.8, -.96) {$x_{\text{tr}}$};
    \end{tikzpicture}
    }
    \end{tabular}
\end{tabular}
  \caption{The computational domain $\Omega$ (a), together with the two
    different current source configurations (b, c) used in the numerical
    computations. The energy transmission ratio is computed by
    measuring $\vec{E}_T^2$ at $x=x_{\text{tr}}$ and $x=x_{\text{re}}$.
  }
  \label{fig:computational-domain}
\end{figure}
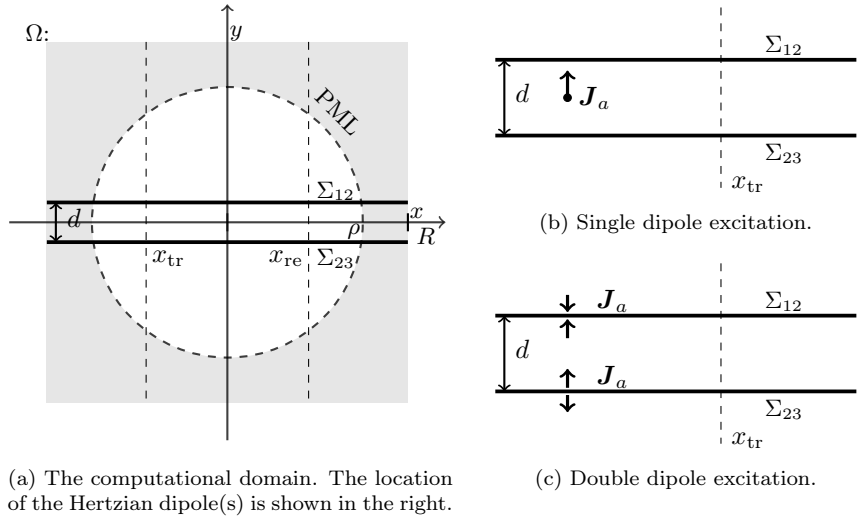

In this paper we will study a prototypical geometry consisting of two flat,
conducting layers in a square domain. The two layers are arranged parallel
to each other with distance $d$ apart; see
Figure~\ref{fig:computational-domain}. This prototypical geometry is
motivated by proposed waveguide configurations that include for example
concentrically arranged carbon nanotubes as integral part of their
design~\cite{prb2009} (see Figure~\ref{fig:MWCNT}. Even though our waveguide configuration is quite
simple in comparison, we make the claim that due to the dominance of the
SPP interaction of the two layers we actually capture the quantitative
behavior of the two-layer interaction quite well. Our computational
framework has thus the potential of guiding the design of more complicated
waveguide structures in the future.

\begin{figure}[t]
  \centering
\tdplotsetmaincoords{70}{46+90} 
\begin{tikzpicture}[scale=0.8,
                    >=stealth',           
                    tdplot_main_coords   %
                    ]

\newcommand{\rotatedtangentangle}[1]{%
    \path[tdplot_rotated_coords] (1,0,0);
    \pgfgetlastxy{\axisxx}{\axisxy}
    \path[tdplot_rotated_coords] (0,1,0);
    \pgfgetlastxy{\axisyx}{\axisyy}
    \path[tdplot_rotated_coords] (0,0,1);
    \pgfgetlastxy{\axiszx}{\axiszy}
    \pgfmathsetmacro{\rtang}{atan(-\axiszy/\axiszx)+180}
    \pgfmathsetmacro{\angkorr}{atan(\axisyy/\axisyx)/2}

    \pgfmathsetmacro{#1}{\rtang+\angkorr}
    }%
    \draw[thick,->] (-1,8,1.0) -- (-2,8,1.0) node[anchor=north west]{\textbf{z}};
    \draw[thick,->] (-1,8,1.0) -- (-1,9,1.0) node[anchor=north west]{\textbf{y}};
    \draw[thick,->] (-1,8,1.0) -- (-1,8,2.0) node[anchor=south]{\textbf{x}};

    \tdplotsetthetaplanecoords{90} 
    \rotatedtangentangle{\tangent} 
    \coordinate (shift) at (0,0,0);
    \tdplotsetrotatedcoordsorigin{(shift)}
    \begin{scope}[tdplot_rotated_coords]
        \draw[fill=gray]
            (0,0,-1) ++(\tangent:1.0) -- ++(0,0,7) arc (\tangent:\tangent-180:1.0) -- ++(0,0,-7);
        \draw[fill=white] (0,0,-1) circle [radius=1.0];
        \draw[fill=white] (0,0,-1) circle [radius=0.75];
        \draw[fill=white] (0,0,-1) circle [radius=0.5];
        \draw[fill=white] (0,0,-1) circle [radius=0.25];
    \end{scope}
  \end{tikzpicture}
  \caption{Schematic of a prototypical multiwall carbon nanotube.}
  \label{fig:MWCNT}
\end{figure}
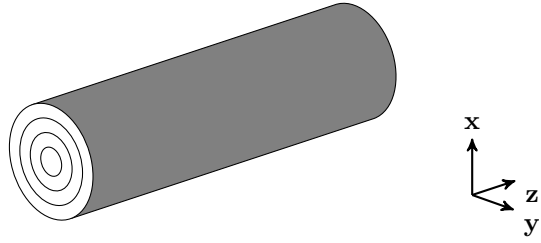

Two different current sources are considered: a single vertical Hertzian
dipole placed at the midpoint of the two sheets (see
Figure~\ref{fig:computational-domain}b); and a double dipole configuration
with one dipole placed directly on each sheet (see
Figure~\ref{fig:computational-domain}c).
The use of such dipole sources in order to excite the desired SPP modes in
the waveguide configuration is again an idealization. However, due to the
large magnitude of the excited SPP modes, the influence of the dipole is
neglibible already after  a short distance from the source. Only the
dominant travelling mode is observed, which enables us to study essentially
``free'' coupled SPP structures in the waveguide.

In the numerical computation we regularize dipoles to a small but finite
thickness. This necessitates a change of sign in the double dipole
configuration when crossing over the interface $\Sigma$. See
Section~\ref{subsec:setup} for details. In the case of the double dipole
excitation, the proximity of the sources to the interface renders the SPP
mode that is excited to dominate \cite{margetis15}. For the single dipole
excitation, however, the strength of the SPP vanishes exponentially with
the distance $d$. For this reason, we will primarily focus on numerical
results obtained with double dipole excitation, 2(b).

The values $x_{\text{tr}}$ and $x_{\text{re}}$ are the transmission and
reception locations, respectively, at which the tangential component of the
energy is to be measured.
In principle, a number of choices for the quantity of interest,
$\mathcal{J}(\vE)$, are possible. In the following, we use a non-linear
quantity of interest given by an \emph{energy transmission ratio}:
\begin{align}
  \mathcal{J}(\vE)=
  \frac{\int_{-R}^R\cos^2\big(\frac{\pi y}{2R}\big)
  |\vE_T|^2(x=x_{\text{re}},y)\dy}
  {\int_{-R}^R\cos^2\big(\frac{\pi
  y}{2R}\big)|\vE_T|^2(x=x_{\text{tr}},y)\dy}.
  \label{eq:quantityofinterest}
\end{align}
The numerator computes the transmitted energy, measured at a vertical strip
located sufficiently far from the source, and the denominator calculates
the received energy. The integrands in $\mathcal{J}({\vE})$ are modified by
weight functions that localize the integral to a vertical strip where we
measure the field intensity for transmission and reception. The choice
\eqref{eq:quantityofinterest} for the quantity of interest leads to a
localized right-hand side $\mathcal{J}$ of the dual problem that is
sensitive to the highly oscillatory SPPs associated with the electric
field, $\vE$.

\emph{Remark.} The functional $\mathcal{J}(\vE)$ is continuously
differentiable as long as the denominator is different from zero. This is
indeed the case for the solution $\vE$ and all approximations $\vE_h$ for
our choice of geometry and dipole excitation.


\subsection{Perfectly Matched Layer}

A perfectly matched layer (PML) is a truncation procedure motivated from
\emph{electromagnetic} scattering problem in the time domain. The
underlying idea of a PML is to surround the computational domain with an
artificial \emph{sponge} layer such that all outgoing electromagnetic waves
decay exponentially with minimal artificial reflection~\cite{berenger94,
chew94,monk03}.

As outlined in~\cite{maier17a, monk03}, we carry out a change of
coordinates from the computational domain with real-valued coordinates to a
domain with complex-valued coordinates.
Projecting back to the real coordinates yields again system
\eqref{eq:timeharmonicmaxwell} with \eqref{eq:jumpcondition}, but with
\emph{modified} material parameters $(\er,\muri,\ssr)$ \emph{inside} the
PML. We refer the reader to \cite{maier17a} for details.

The PML can be implemented by suitably replacing $(\er,\muri,\ssr)$
within the PML. For a spherical absorption layer we define the
matrices
\begin{gather}
  A=T^{-1}_{\vec e_x\vec e_r}\text{diag}\,
  \Big(\frac{1}{\bar d^2},\frac{1}{d\bar d},\frac{1}{d\bar d}\Big)
  T_{\vec e_x\vec e_r},
  \quad
  B=T^{-1}_{\vec e_x\vec e_r}\text{diag}\,\big(d,\bar d,\bar d\big)
  T_{\vec e_x\vec e_r},
  \\\notag
  C=T^{-1}_{\vec e_x\vec e_r}\text{diag}\,\Big(\frac{1}{\bar d},\frac{1}{\bar d},
  \frac{1}{d}\Big)
  T_{\vec e_x\vec e_r},
\end{gather}
\begin{align}
  \label{eq:d}
  d=1+i\,s(r), \quad
  \bar d=1+i/r\int_\rho^r s(\tau)\,\text{d}\tau.
\end{align}
Here, $r$ denotes the distance to the origin, $s(\tau)$ is an appropriate
nonnegative scaling function that will be defined later, $T_{\vec e_x\vec
e_r}$ is the rotation matrix that rotates $\vec e_r$ onto $\vec e_x$. The
material parameters are hence transformed inside the PML as follows:
\begin{align}
  \begin{cases}
    \begin{aligned}
      &\muri
      &\longrightarrow&\quad
      B\muri A,
      \\[0.1em]
      &\er
      &\longrightarrow&\quad
      A^{-1}\er B^{-1},
      \\[0.1em]
      &\ssr
      &\longrightarrow&\quad
      C^{-1}\ssr B^{-1}.
    \end{aligned}
  \end{cases}
  \label{eq:sphericalpmlcoefficients}
\end{align}


\subsection{A posteriori error estimation and local refinement}
\label{subsec:adaptive}

One of the computational challenges of our problem is the need for a much
finer mesh refinement near the interfaces $\Sigma_{12}$ and $\Sigma_{23}$
in order to resolve all small scale SPP structures. We discuss now an
efficient adaptive refinement scheme utilizing an a posteriori error
estimator based on the \emph{dual weighted residual} (DWR)
method~\cite{becker2001}.

Consider the following \emph{dual} problem: Find a solution $\vZ\in\vec
H(\text{curl};\Omega)$ such that
\begin{multline}
  \int_\Omega \left[(\muri\nabla\times\vp)\cdot(\nabla\times\bar\vZ)
  -\er\vp\cdot\bar\vZ\right]\dx
  \\
  -\int_\Sigma\ssr\vp_T\cdot\bar\vZ\dox
  +\;
  \int_{\p\Omega}\sqrt{\muri\er}\vp\cdot\bar\vZ\dox
  = D_E\J(E)[\vp],
  \label{eq:dualproblem}
\end{multline}
for all $\vp\in\vec X(\Omega)$, where $\J(E)$ is a \emph{quantity of
interest} mapping
\begin{align}
  \J:\vec H(\text{curl};\Omega) \to \mathbb{C}.
\end{align}
The dual solution $\vZ$ \emph{encodes} how the target error quantity
depends on local properties of the data \cite{becker2001}. Next, we define
\emph{local error indicators} with the help of the solutions $\vE$ and
$\vZ$ of the primal problem \eqref{eq:variationalformulation} and dual
problem \eqref{eq:dualproblem}, respectively \cite{maier17a},
\cite[Prop.\,2.1]{becker2001}:
\begin{gather}
  \big|\mathcal{J}(\vE)-\mathcal{J}(\vE_h)\big|
  \le \sum_{Q\in\,\TTH}\eta_Q
  +R,\quad\text{with}
  \quad
  \eta_Q:= \frac 12\,\Big| \rho_Q + \rho^\ast_Q \Big|.
  \label{eq:localindicator}
\end{gather}
Here, $\rho_Q$ and $\rho^\ast_Q$ denote the primal and dual cell-wise
residual, respectively, associated with variational equations
(\ref{eq:variationalformulation}) and (\ref{eq:dualproblem}):
\begin{align}
  \label{eq:rhoQ}
  \rho_Q &=
  i\int_\Omega\vJa\cdot\big((\bar\vZ-\bar\vZ_h)\chi_Q\big)\dx
  -A\big(\vE_h,(\vZ-\vZ_h)\chi_Q\big),
  \\
  \label{eq:rhoastQ}
  \rho^\ast_Q &=
  \text{D}_{\vE}\mathcal{J}(\vE_h)[(\vE-\vE_h)\chi_Q]
  -A\big((\vE-\vE_h)\chi_Q,\vZ_h\big),
\end{align}
where $A(\,.\,)$ is given in \eqref{eq:bilinear_form}. Here, $\chi_Q$
denotes the indicator function associated to $Q$, that is, $\chi_Q(\vx)$ is
1 inside the cell $Q$, and 0 otherwise. The local error indicator $\eta_Q$
given by \eqref{eq:localindicator} can now be approximated and used in a
local refinement strategy \cite{becker2001}.

\emph{Remark.} The remainder term $R$ is cubic in the error $\|E-E_h\|$ and
can therefore generally be neglected \cite{becker2001}. More precisely, for
our particular choice of quantity of interest \eqref{eq:quantityofinterest}
a lengthy calculation reveals
\begin{align*}
  |R|&=\Big|\frac12
  \int_0^1\text{D}_{\vE}^3\mathcal{J}(\vE_H+s(\vE-\vE_H))[\vE-\vE_h]^3s(s-1)
  \text{d}s\Big|
  \\
  &\lesssim
  \Big(\frac1R\int_{-R}^R\cos^2\big(\frac{\pi
  y}{2R}\big)|\vE_T|^2(x=x_{\text{tr}},y)\dy\Big)^{-3}
  \big\|\vE-\vE_h\big\|^3,
\end{align*}
provided that the numerator in \eqref{eq:quantityofinterest} is smaller than
the denominator (which is true for our choice of geometry). Given the fact
that we place the measurement position $x_{\text{tr}}$ close to the source
we conclude that $|R|$ is well controlled and small in our case.

Our goal is an optimal local refinement for the numerical simulation of
energy propagation of the SPPs at the location of our choosing.
Consequently, the \emph{weight} $\vZ-\vZ_h$ in residual \eqref{eq:rhoQ} is
generally large near the interface and at points where the influence of the
solution on quantity \eqref{eq:quantityofinterest} is high.

In practice, the numerical evaluation of (\ref{eq:rhoQ}) and
(\ref{eq:rhoastQ}) is typically done with the use of a higher-order
approximation for the dual solution $\vZ$ and the primal solution $\vE$.
However, such a calculation of a higher-order approximation is
computationally costly. We therefore use a patch-wise projection
$\pi_{2H}^{(2)}\vZ_h$ to a higher-order space on a coarser mesh
level~\cite{richter2006}:
\begin{align}
   \vZ-\vZ_h \approx \pi_{2H}^{(2)}\vZ_h - \vZ_h,
   \quad
   \vE-\vE_h \approx \pi_{2H}^{(2)}\vE_h - \vE_h.
\end{align}


\section{Direct numerical simulation}
\label{sec:dns}

In this section we present computational results for the two-layer system
that was introduced above. We demonstrate numerically how the (effective)
wavenumber of SPP structures depends on the interlayer distance $d$ and
investigate the functional relationship of the energy transmission ratio to
the interlayer distance $d$. We determine the optimal spacing, which will
later be used to compare against the analytical findings from
Section~\ref{sec:analytical}. We validate our local refinement strategy by
comparing the convergence rates with uniform refinement, and demonstrate
the effectiveness of the numerical tools discussed in
Section~\ref{sec:numerics}. All numerical computations are carried out with
the finite element library deal.II~\citep{dealii85}.


\subsection{Setup and discretization parameters}
\label{subsec:setup}

We consider a vertical electric dipole positioned at $\vec a_1 = (-0.7,0)$
(for single dipole excitation), and at $\vec a_{2/3} = (-0.7, \pm
d/2)$ (for the double dipole excitation). The current density $\vJa$
is thus given by
\begin{align}
  \vJa =
  \binom{0}{J_0}
  \delta(\vec x - \vec a_1),\text{ and}
  \qquad
  \vJa =
  \binom{0}{J_0}
  \big(\delta_2(\vx - \vec a_2) + \delta_3(\vx-\vec a_3)\big),
\end{align}
for single, or double dipole excitation, respectively. We use two values
for the surface conductivities,
$\sigma_{r,12}^\Sigma=\sigma_{r,23}^\Sigma=\ssr$,
\begin{align*}
  \ssr=0.002 + 0.2i,\text{ and } 0.002 + 0.15i,
\end{align*}
that are within realistic parameter ranges \cite{maier17a}. The
computational domain, $\Omega$, is chosen to be a square with edge length
$4$. A spherical PML is enforced for $\rho > 1.6$. Following
\cite{monk03,maier17a}, we regularize the Dirac deltas in the current
density as follows,
\begin{align*}
  \delta_i(\vx-\vec a_i)
  \;\approx\;
  \frac{\sgn\big((y-d/2)(-y-d/2)\big)\cos^2\big(\pi/(2\,r_d)\,\|(\vx-\vec
  a_i)\|_{2}\big)}
  {\big(\frac{\pi}{2}-\frac{2}{\pi}\big)\,r_d^2},
\end{align*}
for $\|\vx-\vec a_i\|_{2} < r_d$, and $0$ otherwise. The signum function
ensures that the regularized dipole changes sign whenever the
regularization crosses the conducting layer $\Sigma_{12}$, or
$\Sigma_{23}$. We choose a fixed value of $r_d = 10\cdot 2^{-12}$
throughout the paper. We set the position at which we evaluate
\eqref{eq:quantityofinterest} to $x_{\text{tr}} = -0.65$ and $x_{\text{re}}
= 0.75$. This choice maximizes the distance $|x_{\text{re}}-x_{\text{tr}}|$
while ensuring that evaluation points are sufficiently far away from the
regularized dipole sources and the PML.

We use the following scaling function $s(\rho)$ for the PML
\cite{maier17a}
\begin{align}
  s(\rho) = s_0\,\frac{(\rho - 0.8 R)^2}{(R-0.8 R)^2},
\end{align}
and set the free parameter to $s_0=0.05$ in our computations.


\subsection{Validation of local refinement strategy}

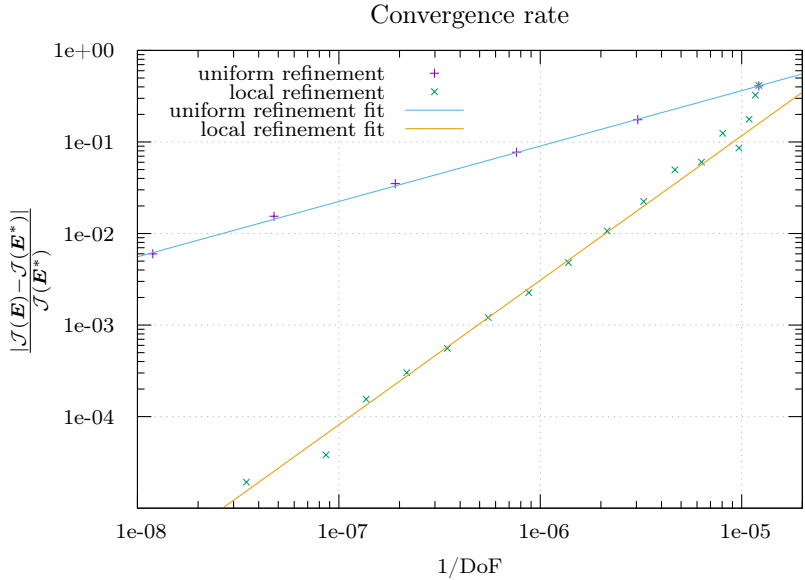
\begin{figure}[tb]
  \centering
  \input{bob-local-uniform.tex}
  \caption{Convergence of the energy tranmission ratio for uniform and local
    refinement for $d = 0.20$. The relative error of the transmission ratios
    obtained with both refinement strategies, respectively, are plotted against
    the inverse of the number of degrees of freedom. We observe a
    convergence order of about $c\approx0.6$ for uniform refinement and
    $c\approx1.6$ for local refinement. The reference value
    $\mathcal{J}{\vE^*}$ was obtained by taking a weighted average of the
    asymptotic transmission ratios,
    $c_\text{local}$ and $c_\text{uniform}$.}
    \label{fig:convergence}
\end{figure}

We validate our numerical framework by comparing values for the
quantity of interest \eqref{eq:quantityofinterest} obtained by
extrapolating numerical values under uniform, and under local refinement.
The computations were performed for $d = 0.20$.
The results are shown in Figure~\ref{fig:convergence}.
The data is fitted to the curve $f(x) = a + b\,x^c$ and extrapolated. For
local refinement, the fit parameters we obtain are $a_{\text{local}} =
0.205333,$ $b_{\text{local}} = 1.25229\times10^8,$ $c_{\text{local}} = 1.57811$,
whereas for uniform refinement we get $a_{\text{unif}} = 0.206216,$
$b_{\text{unif}} = 85.8656,$ $c_{\text{unif}} = 0.612326$. As expected
\cite{maier17a}, we obtain a much faster convergence rate
($c\approx 1.6$) in the quantity of interest for local refinement as opposed
to uniform refinement ($c\approx0.6$). We conclude that our computation of
the energy transmission ratio with $\mathcal{J}(\vE)\approx
a_{\text{local}}$ is reliable within 1\%.


\subsection{Optimal spacing}

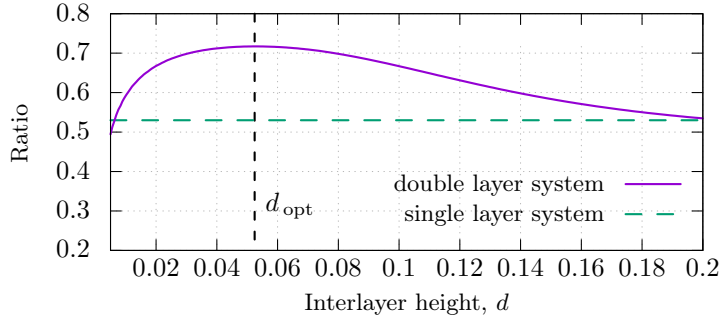
\begin{figure}[tb]
  \centering
  \input{dual-dual.tex}
  \caption{Energy transmission ratio as a function of interlayer
      spacing $d$ computed for the case $\sigma=0.002+0.2i$. The
      computed optimal spacing for a maximal energy transmission ratio is at
      $d_{\,\text{opt}}=0.05245$. The dashed horizontal line is the
      energy transmission ratio computed for the control case of a
      single-layer sheet. For large $d$, the energy transmission ratio of
      the two-layer system approaches the value for the single-layer
      control case asymptotically.}
  \label{fig:ratio-comparison}
\end{figure}
\begin{figure}[tb]
  \centering
  \subfloat[Single-layer control case.]{
    \includegraphics[width=6.0cm]{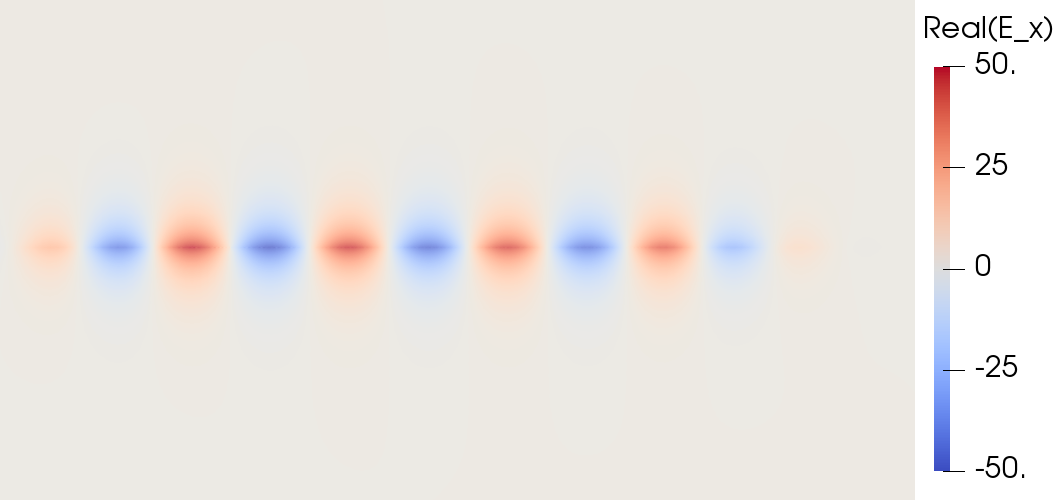}
  }
  \subfloat[Two-layer configuration with $d=d_{\,\text{opt}}\color{black}$]{
    \includegraphics[width=6.0cm]{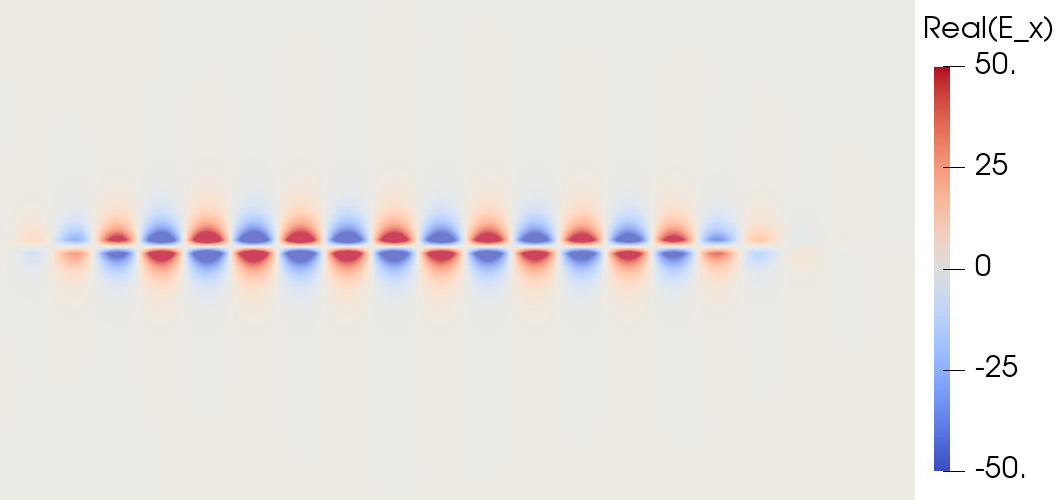}
  }
  \\
  \subfloat[Two-layer configuration with $d\approx0.18$.]{
    \includegraphics[width=6.0cm]{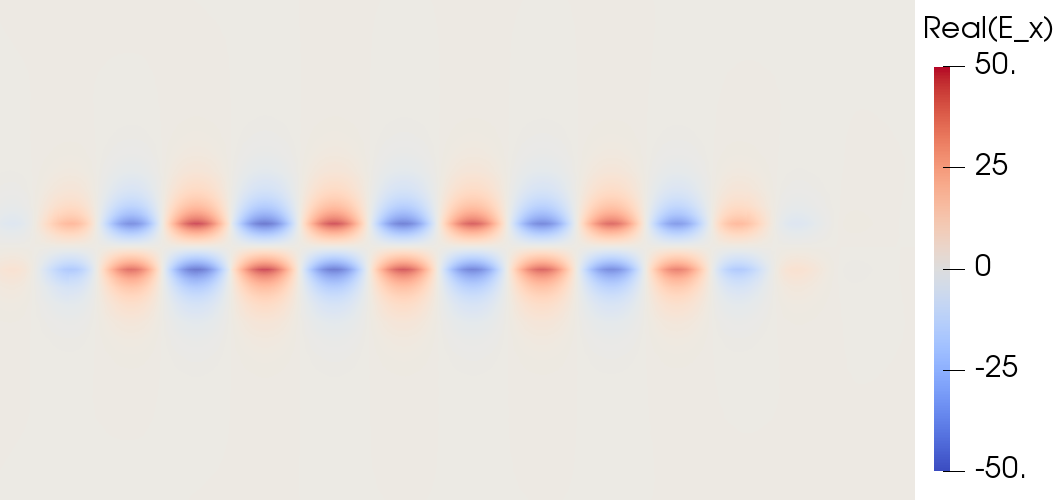}
  }
  \caption{Plasmons observed in different configurations, with $\ssr =
  0.002 +0.2i$. The wavenumber of the excited SPPs in the two-layer
  case (b) with $d=d_{\,\text{opt}}\approx0.05245$ is roughly twice as
  large as the one obtained in the single-layer control case (a), or the
  two-layer configuration (b) with large spacing $d\approx0.18$. The strong
  interlayer coupling for $d=d_{\,\text{opt}}$ in (b) results in a much
  higher SPP amplitude.}
  \label{fig:plasmon}
\end{figure}

Next, we preform a parameter study of the energy transmition ratio for
varying interlayer spacing $d$ ranging from $d_{\text{min}}=4\cdot10^{-12}$
to $d_{\text{max}}=0.2$, where $d_{max}$ corresponds to about 1/3 of the
single-layer SPP wavelength, $2\pi/\text{Re}(k_{m,r})$, or 1/30 of the free
space wavelength~\cite{maier17a}. The (interpolated) results are shown in
Figure~\ref{fig:ratio-comparison}. We used a releatively coarse initial
mesh for all computations with around 20 thousand degrees of freedom. After
12 local refinement cycles using the adaptive refinement procedure outlined
in Section~\ref{subsec:adaptive}, we reached roughly 2 million degrees of
freedom on the finest mesh. We make a qualitative comparison of three
representative cases: Figure~\ref{fig:plasmon} shows the real part of the
computed scattered electric field  in $x$-direction,
$\text{Re}(E_x^\text{sc})$, for SPPs on a single-layer system, and a
two-layer system with $d=d_{\,\text{opt}}=0.05245$ and $d=0.1805$,
respectively. In the case of optimal spacing, $d\,=\,d_{\,\text{opt}}$, the
wavenumber of the excited SPPs is roughly twice as large as the one
obtained for the single-layer case, cf. Figures~\ref{fig:plasmon}a
and~\ref{fig:plasmon}b. For large enough $d$, for example $d\approx\,0.18$,
we observe that the wavenumber of the excited SPP approaches the
single-layer case, cf. Figures~\ref{fig:plasmon}a and~\ref{fig:plasmon}c.
This indicates that the two-layer system is converging to the single-layer
setting, analogous to the behavior observed for the energy transmission
ratio.


\section{Analytic solution and validation}
\label{sec:analytical}

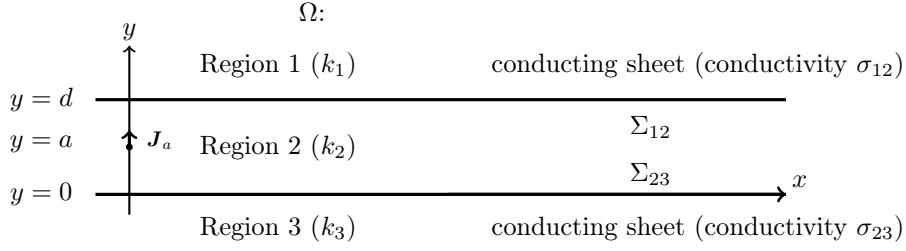
\begin{figure}[t]
  \centering
    \begin{tikzpicture}[
      scale=0.9,
      extended/.style={shorten >=-#1, shorten <=-#1},
      extended/.default=1cm]
      \node at (-2.0, 2.0) {$\Omega$:};
      \path [->, opacity=2.00, very thick, draw] (-0, -0.7) -- (5.00, -0.7);
      \path [->, opacity=1.00, thick, draw] (-4.7, -1.0) -- (-4.7, 1.5);
      \node at (5.2, -0.5) {$x$};
      \node at (-4.7, 1.7) {$y$};
      \path [very thick,draw] (-5.2, 0.7) -- (5.0, 0.7);
      \path [very thick,draw] (-5.2, -0.7) -- (5.0, -0.7);
      \node at (-2.5, 1.20) {Region 1 ($k_1$)};
      \node at (3.7,1.20) {conducting sheet (conductivity $\sigma_{12}$)};
      \node at (-2.5, 0.00) {Region 2 ($k_2$)};
      \node at (-2.5, -1.20) {Region 3 ($k_3$)};
      \node at (3.7,-1.20) {conducting sheet (conductivity $\sigma_{23}$)};
      \node at (-6.0, -.7) {$y=0$};
      \node at (-6.0, 0.05) {$y=a$};
      \node at (-6.0, 0.7) {$y=d$};
      \node at (3.0,0.3) {$\Sigma_{12}$};
      \node at (3.0,-0.4) {$\Sigma_{23}$};
      \path [very thick, ->, draw] (-4.7, 0.0) -- (-4.7, 0.25);
      \node[scale=0.8] at (-4.25, 0.1) {$\vJa$};
      \draw[fill] (-4.7,0.0) circle (1.2pt);
    \end{tikzpicture}
    \vspace{1em}
  \caption{Schematic of a vertical electric dipole at a distance $a$ from
  conducting sheet $\Sigma_{23}$ in 2D. The dipole has a current density
  $\vJa=J_0\delta(x)\delta(y-a)\vec e_y$. The
  bottom sheet lies on $y=0$ and the top sheet lies on $y=d$. Sheets
  separate the space into region 1 ($\{y>d\}$) with wavenumber $k_1$,
  region 2 ($\{0<y<d\}$) with wavenumber $k_2$, and region 3
  $(\{y<0\})$ with wavenumber $k_3$. Each sheet is prescribed with a
  surface conductivity $\sigma_{12}$ and $\sigma_{23}$,
  respectively.
  }
  \label{fig:layeredmedium}
\end{figure}
In this section, we derive an analytic solution for
\eqref{eq:timeharmonicmaxwell} and \eqref{eq:jumpcondition} for an
(idealized) infinite two-layer system with single dipole excitation; see
Figure~\ref{fig:computational-domain}b and Figure~\ref{fig:layeredmedium}).
We identify the limiting behavior for the case of large interlayer spacing
$d$, compute the effective wavenumber of the dominant SPP mode and validate
our numerical findings with these results.

For better readability and ease of comparison \cite{margetis15}, we revert
back from the rescaled version~\eqref{eq:2ndorderequationrescaled} to the
original form of Maxwell's equations \eqref{eq:timeharmonicmaxwell}. Let us
consider two planar sheets in $\mathbb{R}^2$ situated at $y=0$ and $y=d$,
respectively; see Figure~\ref{fig:layeredmedium}. The conducting sheets
separate $\mathbb{R}^2$ into three regions: Region 1 $(\{y>d\})$ has
wavenumber $k_1$ and shares the boundary with region 2 $(\{0<y<d\})$, whose
wavenumber is given by $k_2$. Region 3 $(\{y<0\})$ shall have wavenumber
$k_3$, where $k_j^2=\omega^2\e_j\mu$. Here, $\e_j$ denotes a complex-valued
permittivity $(j=1,2,3)$. Let a vertical electric Hertzian dipole be
positioned at $(0,a)$ in between the interfaces, viz.,
$\vJa=\delta(x)\delta(y-a)\vec e_y$. Define the Fourier transform,
$\hat{\vec{F}}(\xi,y)$ of the vector-valued fields $(\vec{F}\equiv\vB,\vE)$
through the integral formula
\begin{align}
  \vec{F}(x,y) = \frac1{2\pi}\int_\mathbb{R}\,d\xi
  \;\hat{\vec{F}}(\xi,y) e^{i\xi x}.
\end{align}
Applying the Fourier transform to Maxwell's equations
\eqref{eq:timeharmonicmaxwell} gives
\begin{align}
  \begin{cases}
    \begin{aligned}
      & -i\xi\hat{E}_{jy} + \frac\p{\p y}\hat{E}_{jx}
        = -i\omega\hat{B}_{jz}, \\
      & -\frac{\p}{\p y}\hat{B}_{jz} = \frac{ik_j^2}{\omega}\hat{E}_{jx}, \\
      & -i\xi\hat{B}_{jz} =
      -\frac{ik_j^2}{\omega}\hat{E}_{jy}+\mu\delta(y).
      \label{eq:fouriertransform_maxwell}
    \end{aligned}
  \end{cases}
\end{align}
A number of elementary algebraic manipulations of
(\ref{eq:fouriertransform_maxwell}) yield the differential equation
\begin{align}
  \left(\frac{\p^2}{\p y^2} + \beta_j^2 \right)\hat{B}_{jz} =
    -i\xi\mu\delta(y),
\end{align}
where we have set $\beta_j^2 = k_j^2-\xi^2$. We now make the following
solution ansatz for the magnetic field obeying the Sommerfeld radiation
condition \cite{margetis15}:
\begin{align}
  \label{eq:ansatz}
  \hat B_{jz} (\xi,y) = \left\{\begin{array}{lll}
  a_1 e^{i\beta_1 y}, & y > d, \\
  a_2e^{i\beta_2 y} + b_2e^{-i\beta_2 y} - \frac{\xi\mu}{2\beta_2}
  e^{i\beta_2|y-a|},
  & 0<y<d, \\
  b_3e^{-i\beta_3 y}, & y < 0. \end{array}\right.
\end{align}
The remaining electric field components can be derived from the relations
\begin{align*}
  \hat{E}_{jx} & = \frac{i\omega}{k_j^2}\frac{\p}{\p y}
    \hat{B}_{jz}, \\
  \hat{E}_{jy} & = \frac{\omega}{k_j^2\xi}\left(\xi^2\hat{B}_{jz}
    - i\xi\mu\delta(y)\right).
\end{align*}
Next, we determine closed expressions for the coefficients in
\eqref{eq:ansatz} by matching with boundary
conditions~\eqref{eq:jumpcondition} on each interface:
\begin{align*}
    & a_1 = \frac{\beta_2k_1^2}{\beta_1k_2^2}\bigg[a_2e^{i\beta_2d}
      -b_2e^{-i\beta_2d}-\frac{\xi\mu}{2\beta_2}e^{i\beta_2(d-a)}\bigg]
      e^{-i\beta_1d}, \\
    & a_2 = -\frac{\xi\mu}{2\beta_2} \frac{R_{23}
      (e^{i\beta_2a)}+R_{12}e^{i\beta_2(2d-a)})}
      {1-R_{12}R_{23}e^{2i\beta_2d}},
\end{align*}
\begin{align*}
    & b_2 = -\frac{\xi\mu}{2\beta_2} \frac{R_{12}e^{i\beta_2 d}
      (e^{i\beta_2(d-a)}+R_{23}e^{i\beta_2(d+a)})}
      {1-R_{12}R_{23}e^{2i\beta_2d}},\\
    & b_3 = -\frac{\beta_2k_3^2}{\beta_3k_2^2}\bigg[a_2
      -b_2+\frac{\xi\mu}{2\beta_2}e^{i\beta_2a}\bigg].
\end{align*}
Here, the constants $R_{ij,m}$ are given by
\begin{align}
  R_{ij} & = \frac{\beta_i k_j^2-\beta_j k_i^2+\omega\mu\sigma_{ij}\beta_i\beta_j}
  {\beta_ik_j^2+\beta_jk_i^2+\omega\mu\sigma_{ij}\beta_i\beta_j}.
\end{align}
By substituting back into \eqref{eq:ansatz} and undoing the Fourier
transform, all field components of $\vE$ and $\vB$ can be expressed as
analytic integrals. In particular, we are interested in $E_{2x}(x,y)$, the
electric field component in $x$-direction between the two-layers:
\begin{multline}
  \label{eq:e2x}
  E_{2x}(x,y)=\frac{\omega\mu}{4\pi k_2^2}\int_{-\infty}^\infty\,d\xi
    \;\ \xi \bigg[\frac{R_{23}
    (e^{i\beta_2a}+R_{12}e^{i\beta_2(2d-a)})}
    {1-R_{12}R_{23}e^{2i\beta_2d}}e^{i\beta_2y}\\
    -\frac{R_{12}e^{2i\beta_2d}(e^{-i\beta_2a}+R_{23}e^{i\beta_2a})}
    {1-R_{12}R_{23}e^{2i\beta_2d}}e^{-i\beta_2y}
    +\sgn(y-a)e^{i\beta_2|y-a|}
    \bigg]e^{i\xi x}.
\end{multline}


\subsection{Approximation of the pole contribution}

Next, we obtain the scattered electric field in $x$-direction,
$E_{2x}^\text{sc}$, observed at $y=0$ by subtracting the incident field
\vspace{-0.75em}
\begin{align*}
  \frac{\omega\mu}{4\pi k_2^2}\int_{-\infty}^\infty\,d\xi
  \;\xi\,\sgn(y-a)e^{i\beta_2|y-a|}e^{i\xi x}.
\end{align*}
from \eqref{eq:e2x}. Some additional minor rearrangement yields
\begin{multline}
  E_{2x}^\text{sc}(x,0) =
  \frac{\omega\mu}{4\pi k_2^2}\int_{-\infty}^\infty
  \,d\xi\;\ \xi \bigg[
  \frac{R_{23}e^{i\beta_2 a}}{1-R_{12}R_{23}e^{2i\beta_2 d}}
  -\frac{R_{12}e^{i\beta_2(2d-a)}}{1-R_{12}R_{23}e^{2i\beta_2 d}}\\
  -\frac{R_{12}R_{23}e^{2i\beta_2 d}(e^{i\beta_2 a}-e^{-i\beta_2 a})}
  {1-R_{12}R_{23}e^{2i\beta_2 d}}
  \bigg]e^{i\xi x}
  \;=:\;(\RN{1})+(\RN{2})+(\RN{3}).
  \label{eq:scatteredsolution}
\end{multline}
Each term of the integrand contains SPP contributions stemming from
different conducting sheets: Term (I) and (II) arise from the SPP
situated at $\Sigma_{23}$ and $\Sigma_{12}$, respectively. Term (III) is a
\emph{mixed} term due to the interlayer coupling of SPPs.

We now discuss the role of simple poles in the evaluation of integral
\eqref{eq:scatteredsolution}. For the sake of simplicity, let us now assume
that $k\equiv k_1=k_2=k_3$ and
$\sigma\equiv\sigma_{23}=\sigma_{12}$. Thus, $R_{12} =
R_{23}\equiv R$ and $\beta_j\equiv\beta$, where
\begin{align}
  R := \frac{\omega\mu\sigma\beta}{(2k^2+\omega\mu\sigma\beta)},
  \qquad
  \beta^2:=k^2-\xi^2.
\end{align}
Waveguide modes correspond to \emph{single poles} of the integrand in
integral \eqref{eq:scatteredsolution} \cite{margetis15}. Inspecting
\eqref{eq:scatteredsolution} we see that these are exactly given by the
condition $Re^{i\beta d}=\pm1$. The solution for the branch with the plus
sign recovers even waveguide modes, and, correspondingly, the minus sign
recovers odd waveguide modes~\cite{brueck00}.

For $d$ sufficiently small, there is only a single dominant mode. We
analyze this case further. The common prefactor of term (I) and (II) is
given by
\begin{align}
  \begin{aligned}
    \frac{R}{1-R^2e^{2i\beta d}} & =
      \frac{(2k^2+\omega\mu\sigma\beta)\omega\mu\sigma\beta}
      {(2k^2+\omega\mu\sigma\beta(1-e^{i\beta d}))
      (2k^2+\omega\mu\sigma\beta(1+e^{i\beta d}))}.
  \end{aligned}
  \label{eq:prefactor}
\end{align}
The TM surface plasmon corresponds to the residue contribution to the
electromagnetic field from the pole $\xi=k_m^B$, where $k_m^B$ is a
solution of the transcendental relationship for the mode,
$2k^2+\omega\mu\sigma\beta(1-e^{i\beta d})=0$~\cite{margetis15}.
Now,
\begin{align}
  \begin{aligned}
    2k^2+\omega\mu\sigma&\beta(1-e^{i\beta d})\approx
     \\
     &-(\xi-k_m^B) (id)k_m^B\omega\mu\sigma
     \bigg(e^{i\beta_p d} -
     \frac{1-e^{i\beta_p d}}{id\beta_p}\bigg).
  \end{aligned}
\end{align}
Here, the subscript $p$ denotes evaluation at the pole.
Each of $\RN{1},\RN{2},\RN{3}$ consists of the branch-cut contribution
and the pole contribution. We omit the discussion of the branch-cut in
this paper and focus only on the simple pole, $\xi=k_m^B$. This is because
for an infinite conducting sheet, the SPP is identified with the part
of the electromagnetic field equal to the contribution to the Fourier integrals
of the simple pole that solves the above transcendental relationship.
For a more thorough discussion on the branch-cut and its computation, we refer readers to~\cite{maier17a}.
By the residue theorem,
\begin{align}
  \label{eq:firstterm}
  (\RN{1})&=\frac{\omega\mu}{4\pi k^2}\int_{-\infty}^\infty \,d\xi\;\ \xi
  \frac{Re^{i\beta a}e^{i\xi x}}{1-R^2 e^{2i\beta d}} = (\RN{1})^p + (\RN{1})^{b.c.};
  \\\notag
  (\RN{1})^p&\approx-\frac{i\omega\mu\beta_p^2}{2k^2}\frac{2k^2+\omega\mu\sigma\beta_p}
  {2k^2+\omega\mu\sigma\beta_p(1+e^{i\beta_pd})}\frac{e^{i(k_m^Bx+\beta_pa)}}
  {1-e^{i\beta_pd}(1+i\beta_pd)}.
  \intertext{And similarly for the second integrand term,}
  \label{eq:secondterm}
  (\RN{2})&=-\frac{\omega\mu}{4\pi k^2}\int_{-\infty}^\infty \,d\xi\;\ \xi
  \frac{Re^{i\beta(2d-a)}e^{i\xi x}}{1-R^2 e^{2i\beta d}} = (\RN{2})^p + (\RN{2})^{b.c.};
  \\\notag
  (\RN{2})^p&\approx\frac{i\omega\mu\beta_p^2}{2k^2}\frac{2k^2+\omega\mu\sigma\beta_p}
  {2k^2+\omega\mu\sigma\beta_p(1+e^{i\beta_pd})}\frac{e^{i(k_m^Bx+\beta_p(2d-a))}}
  {1-e^{i\beta_pd}(1+i\beta_pd)}.
  \intertext{The interlayer pole contribution is calculated in the same
    fashion.}
  \label{eq:thirdterm}
  (\RN{3})&=-\frac{\omega\mu}{4\pi k^2}\int_{-\infty}^\infty \,d\xi\;\ \xi
  \frac{R^2e^{2i\beta d}(e^{i\beta a}-e^{-i\beta a})}{1-R^2e^{2i\beta d}}
  e^{i\xi x} = (\RN{3})^p + (\RN{3})^{b.c.};
  \\\notag
  (\RN{3})^p&\approx\frac{\omega\mu\beta_p^3}{k^2}\frac{\omega\mu\sigma\sin(\beta_pa)}
  {2k^2+\omega\mu\sigma\beta_p(1+e^{i\beta_pd})}\frac{e^{i(k_m^Bx+2\beta_pd)}}
  {1-e^{i\beta_pd}(1+i\beta_pd)}.
\end{align}
%


\subsection{Limiting behavior and effective SPP wavenumber}

\begin{figure}[tb]
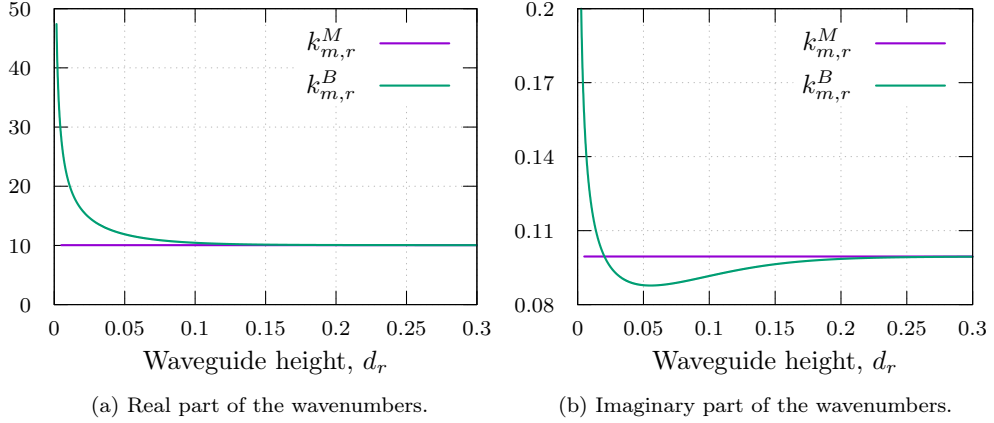

  \centering
  \subfloat[Real part of the wavenumbers.]{\input{k_m-comparison-real.tex}}
  \kern-1.5em
  \subfloat[Imaginary part of the wavenumbers.]{\input{k_m-comparison-imag.tex}}
  \caption{Convergence of the two-layer system wavenumber $k_{m,r}^B$ to
    single-layer wavenumber $k_{m,r}^M$ numerically computed with a root
    solver for \eqref{eq:pole}. The minimum loss is observed at
    $d_r=d_r^{\,\text{opt}}\approx 0.05538$.}
  \label{fig:k_m-convergence}
\end{figure}

For $d\gg1$ and fixed dipole position $a$,  we expect the solution of the
double-layer system to approach the solution of the single-layer system. We
justify this claim by observing that as $d\to\infty$,
\begin{align}
  \begin{aligned}
    (\RN{1})^p \to -\frac{i\omega\mu}{2k^2}\beta_p^2e^{i(\beta_pa+k_m^Bx)} &=
    -\frac{i\omega\mu}{2k^2}\frac{4k^4}{(\omega\mu\sigma)^2}
    e^{i(\beta_pa+k_m^Bx)} \\
    & =
    -2i\omega\mu\bigg(\frac{k}{\omega\mu\sigma}\bigg)^2e^{ik_m^Bx
    -2iak^2/(\omega\mu\sigma)},
  \end{aligned}
\end{align}
which corresponds to the single-layer solution~\cite{maier17a}. Further,
the remaining terms $(\RN{2})_p$ and $(\RN{3})_p$ converge to zero due to
the presence of $e^{i\beta_pd}$. Under the rescaling introduced in
Section~\ref{sec:variational}, equation (\ref{eq:prefactor}) becomes
\begin{align}
  \label{eq:pole}
  2\mur\er+\ssr\beta_{p,r}(1-e^{i\beta_{p,r}d_r})=0,
  \text{ with }\beta_{p,r}=\sqrt{\mur\er-(k^B_{m,r})^2}.
\end{align}
We numerically solve the above via a root finding algorithm and plot the
real and the imaginary part of the wavenumber $k^B_{m,r}$ as a function of
distance $d_r$; see Figure~\ref{fig:k_m-convergence}. For $d_r$ small,
$\text{Im}(k_{m,r}^B)$ is greater than the single-layer counterpart.
However, upon entering a regime  where the interlayer coupling dominates,
i.e., where $k_{m,r}^B$ becomes less lossy than $k_{m,r}^M$, we observe
that there is an optimal distance, $d_r^{\,\text{opt}}\approx 0.05538$, at
which the scattered field solution in the $x$-direction attains its maxium.
For $d_r$ large enough, only the contribution from the bottom interface
remains (term (II)$_p$ and (III)$_p$ vanish), and the wavenumber converges
to that of the single-layer case.


\subsection{Comparison and validation of numerical results}

Finally, we compare the numerical results obtained in Section~\ref{sec:dns}
to the analytical solution derived above.

In particular, we expect to observe that the contribution from the
interlayer coupling term (III)$^p$ of the SPP dominates in the energy
transmission ratio. In this vein, our direct numerical simulation is
performed for the double-dipole excitation on the interfaces so that the
(III)$^p$ dominates \cite{maier17a}; see Section~\ref{sec:dns}.
We postulate that the complex-valued wavenumber $k_{m,r}^B$ given by
\eqref{eq:pole} and associated with the SPP mode (III)$^p$ describes the
effective transmission behavior of the two-layer system.

The optimal distance obtained from the analytical solution is computed to
be $d_r^{\,\text{opt}}\approx 0.05538$; see
Figure~\ref{fig:k_m-convergence}. In order to test the validity of our
numerical method, we compare this value against the optimal distance,
$d_{\,\text{opt}}= 0.05245$, obtained by numerically computed the energy
transmission ratio; see Figure~\ref{fig:ratio-comparison}. Both values are
in very good agreement. We attribute the small discrepancy of both values
to the different current sources that were used.


\section{Conclusion}
\label{sec:conclusion}

In this paper, we extended a variational framework for the numerical
simulation of the SPPs excited by a current-carrying source on an infinite
conducting sheet to the SPPs generated by single/double excitations in a
waveguide configuration. The conducting sheets, e.g. graphene, are modeled
as idealized hypersurfaces that naturally takes into account the jump
condition of the magnetic field.

We demonstrate that the interlayer coupling of the SPPs present in the
two-layer system is responsible for higher confinement and less losses than
the single-layer system. We computed optimal interlayer spacings using two
approaches. First, we compute via finite element simulations with
double-dipole excitation. Second, we derive the pole contribution of the
$x$-directed scattered field solution and solve for the wavenumber of the
SPPs \eqref{eq:pole} numerically. The computed values are in agreement with
one another. The numerical results on the energy transmission ratio are in
very good agreement with analytic results obtained for the SPP mode of the
double-layer structure.

Our numerics admit several generalizations and extensions.
In particular, our variational framework can be readily used without
modification to model any geometric configuration that is meshable by
quadrilaterals. This flexibility enables numerical simulations of curved
waveguides, systems consisting of multiple layers, or complicated optical
devices in the near future.


\section*{Acknowledgments}
We wish to thank Professor Dionisios Margetis for useful
discussions. We acknowledge support by ARO MURI Award W911NF-14-0247.


\end{document}

%% file: bob-local-uniform.tex
\begin{tikzpicture}[gnuplot]
\path (0.000,0.000) rectangle (12.000,8.000);
\gpcolor{color=gp lt color axes}
\gpsetlinetype{gp lt axes}
\gpsetdashtype{gp dt axes}
\gpsetlinewidth{0.50}
\draw[gp path] (2.608,2.203)--(11.447,2.203);
\gpcolor{color=gp lt color border}
\gpsetlinetype{gp lt border}
\gpsetdashtype{gp dt solid}
\gpsetlinewidth{1.00}
\draw[gp path] (2.608,2.203)--(2.788,2.203);
\draw[gp path] (11.447,2.203)--(11.267,2.203);
\node[gp node right,font={\fontsize{8.0pt}{9.6pt}\selectfont}] at (2.424,2.203) {{     1e-04}};
\draw[gp path] (2.608,2.570)--(2.698,2.570);
\draw[gp path] (11.447,2.570)--(11.357,2.570);
\draw[gp path] (2.608,2.784)--(2.698,2.784);
\draw[gp path] (11.447,2.784)--(11.357,2.784);
\draw[gp path] (2.608,2.936)--(2.698,2.936);
\draw[gp path] (11.447,2.936)--(11.357,2.936);
\draw[gp path] (2.608,3.054)--(2.698,3.054);
\draw[gp path] (11.447,3.054)--(11.357,3.054);
\draw[gp path] (2.608,3.151)--(2.698,3.151);
\draw[gp path] (11.447,3.151)--(11.357,3.151);
\draw[gp path] (2.608,3.232)--(2.698,3.232);
\draw[gp path] (11.447,3.232)--(11.357,3.232);
\draw[gp path] (2.608,3.303)--(2.698,3.303);
\draw[gp path] (11.447,3.303)--(11.357,3.303);
\draw[gp path] (2.608,3.365)--(2.698,3.365);
\draw[gp path] (11.447,3.365)--(11.357,3.365);
\gpcolor{color=gp lt color axes}
\gpsetlinetype{gp lt axes}
\gpsetdashtype{gp dt axes}
\gpsetlinewidth{0.50}
\draw[gp path] (2.608,3.421)--(11.447,3.421);
\gpcolor{color=gp lt color border}
\gpsetlinetype{gp lt border}
\gpsetdashtype{gp dt solid}
\gpsetlinewidth{1.00}
\draw[gp path] (2.608,3.421)--(2.788,3.421);
\draw[gp path] (11.447,3.421)--(11.267,3.421);
\node[gp node right,font={\fontsize{8.0pt}{9.6pt}\selectfont}] at (2.424,3.421) {{     1e-03}};
\draw[gp path] (2.608,3.788)--(2.698,3.788);
\draw[gp path] (11.447,3.788)--(11.357,3.788);
\draw[gp path] (2.608,4.002)--(2.698,4.002);
\draw[gp path] (11.447,4.002)--(11.357,4.002);
\draw[gp path] (2.608,4.154)--(2.698,4.154);
\draw[gp path] (11.447,4.154)--(11.357,4.154);
\draw[gp path] (2.608,4.272)--(2.698,4.272);
\draw[gp path] (11.447,4.272)--(11.357,4.272);
\draw[gp path] (2.608,4.369)--(2.698,4.369);
\draw[gp path] (11.447,4.369)--(11.357,4.369);
\draw[gp path] (2.608,4.450)--(2.698,4.450);
\draw[gp path] (11.447,4.450)--(11.357,4.450);
\draw[gp path] (2.608,4.521)--(2.698,4.521);
\draw[gp path] (11.447,4.521)--(11.357,4.521);
\draw[gp path] (2.608,4.583)--(2.698,4.583);
\draw[gp path] (11.447,4.583)--(11.357,4.583);
\gpcolor{color=gp lt color axes}
\gpsetlinetype{gp lt axes}
\gpsetdashtype{gp dt axes}
\gpsetlinewidth{0.50}
\draw[gp path] (2.608,4.639)--(11.447,4.639);
\gpcolor{color=gp lt color border}
\gpsetlinetype{gp lt border}
\gpsetdashtype{gp dt solid}
\gpsetlinewidth{1.00}
\draw[gp path] (2.608,4.639)--(2.788,4.639);
\draw[gp path] (11.447,4.639)--(11.267,4.639);
\node[gp node right,font={\fontsize{8.0pt}{9.6pt}\selectfont}] at (2.424,4.639) {{     1e-02}};
\draw[gp path] (2.608,5.006)--(2.698,5.006);
\draw[gp path] (11.447,5.006)--(11.357,5.006);
\draw[gp path] (2.608,5.220)--(2.698,5.220);
\draw[gp path] (11.447,5.220)--(11.357,5.220);
\draw[gp path] (2.608,5.372)--(2.698,5.372);
\draw[gp path] (11.447,5.372)--(11.357,5.372);
\draw[gp path] (2.608,5.490)--(2.698,5.490);
\draw[gp path] (11.447,5.490)--(11.357,5.490);
\draw[gp path] (2.608,5.587)--(2.698,5.587);
\draw[gp path] (11.447,5.587)--(11.357,5.587);
\draw[gp path] (2.608,5.668)--(2.698,5.668);
\draw[gp path] (11.447,5.668)--(11.357,5.668);
\draw[gp path] (2.608,5.739)--(2.698,5.739);
\draw[gp path] (11.447,5.739)--(11.357,5.739);
\draw[gp path] (2.608,5.801)--(2.698,5.801);
\draw[gp path] (11.447,5.801)--(11.357,5.801);
\gpcolor{color=gp lt color axes}
\gpsetlinetype{gp lt axes}
\gpsetdashtype{gp dt axes}
\gpsetlinewidth{0.50}
\draw[gp path] (2.608,5.857)--(11.447,5.857);
\gpcolor{color=gp lt color border}
\gpsetlinetype{gp lt border}
\gpsetdashtype{gp dt solid}
\gpsetlinewidth{1.00}
\draw[gp path] (2.608,5.857)--(2.788,5.857);
\draw[gp path] (11.447,5.857)--(11.267,5.857);
\node[gp node right,font={\fontsize{8.0pt}{9.6pt}\selectfont}] at (2.424,5.857) {{     1e-01}};
\draw[gp path] (2.608,6.224)--(2.698,6.224);
\draw[gp path] (11.447,6.224)--(11.357,6.224);
\draw[gp path] (2.608,6.438)--(2.698,6.438);
\draw[gp path] (11.447,6.438)--(11.357,6.438);
\draw[gp path] (2.608,6.590)--(2.698,6.590);
\draw[gp path] (11.447,6.590)--(11.357,6.590);
\draw[gp path] (2.608,6.708)--(2.698,6.708);
\draw[gp path] (11.447,6.708)--(11.357,6.708);
\draw[gp path] (2.608,6.805)--(2.698,6.805);
\draw[gp path] (11.447,6.805)--(11.357,6.805);
\draw[gp path] (2.608,6.886)--(2.698,6.886);
\draw[gp path] (11.447,6.886)--(11.357,6.886);
\draw[gp path] (2.608,6.957)--(2.698,6.957);
\draw[gp path] (11.447,6.957)--(11.357,6.957);
\draw[gp path] (2.608,7.019)--(2.698,7.019);
\draw[gp path] (11.447,7.019)--(11.357,7.019);
\gpcolor{color=gp lt color axes}
\gpsetlinetype{gp lt axes}
\gpsetdashtype{gp dt axes}
\gpsetlinewidth{0.50}
\draw[gp path] (2.608,7.075)--(11.447,7.075);
\gpcolor{color=gp lt color border}
\gpsetlinetype{gp lt border}
\gpsetdashtype{gp dt solid}
\gpsetlinewidth{1.00}
\draw[gp path] (2.608,7.075)--(2.788,7.075);
\draw[gp path] (11.447,7.075)--(11.267,7.075);
\node[gp node right,font={\fontsize{8.0pt}{9.6pt}\selectfont}] at (2.424,7.075) {{     1e+00}};
\gpcolor{color=gp lt color axes}
\gpsetlinetype{gp lt axes}
\gpsetdashtype{gp dt axes}
\gpsetlinewidth{0.50}
\draw[gp path] (2.608,0.985)--(2.608,7.075);
\gpcolor{color=gp lt color border}
\gpsetlinetype{gp lt border}
\gpsetdashtype{gp dt solid}
\gpsetlinewidth{1.00}
\draw[gp path] (2.608,0.985)--(2.608,1.165);
\draw[gp path] (2.608,7.075)--(2.608,6.895);
\node[gp node center,font={\fontsize{8.0pt}{9.6pt}\selectfont}] at (2.608,0.677) {{      1e-08}};
\draw[gp path] (3.414,0.985)--(3.414,1.075);
\draw[gp path] (3.414,7.075)--(3.414,6.985);
\draw[gp path] (3.886,0.985)--(3.886,1.075);
\draw[gp path] (3.886,7.075)--(3.886,6.985);
\draw[gp path] (4.220,0.985)--(4.220,1.075);
\draw[gp path] (4.220,7.075)--(4.220,6.985);
\draw[gp path] (4.480,0.985)--(4.480,1.075);
\draw[gp path] (4.480,7.075)--(4.480,6.985);
\draw[gp path] (4.692,0.985)--(4.692,1.075);
\draw[gp path] (4.692,7.075)--(4.692,6.985);
\draw[gp path] (4.871,0.985)--(4.871,1.075);
\draw[gp path] (4.871,7.075)--(4.871,6.985);
\draw[gp path] (5.026,0.985)--(5.026,1.075);
\draw[gp path] (5.026,7.075)--(5.026,6.985);
\draw[gp path] (5.163,0.985)--(5.163,1.075);
\draw[gp path] (5.163,7.075)--(5.163,6.985);
\gpcolor{color=gp lt color axes}
\gpsetlinetype{gp lt axes}
\gpsetdashtype{gp dt axes}
\gpsetlinewidth{0.50}
\draw[gp path] (5.286,0.985)--(5.286,5.911);
\draw[gp path] (5.286,6.895)--(5.286,7.075);
\gpcolor{color=gp lt color border}
\gpsetlinetype{gp lt border}
\gpsetdashtype{gp dt solid}
\gpsetlinewidth{1.00}
\draw[gp path] (5.286,0.985)--(5.286,1.165);
\draw[gp path] (5.286,7.075)--(5.286,6.895);
\node[gp node center,font={\fontsize{8.0pt}{9.6pt}\selectfont}] at (5.286,0.677) {{      1e-07}};
\draw[gp path] (6.092,0.985)--(6.092,1.075);
\draw[gp path] (6.092,7.075)--(6.092,6.985);
\draw[gp path] (6.563,0.985)--(6.563,1.075);
\draw[gp path] (6.563,7.075)--(6.563,6.985);
\draw[gp path] (6.898,0.985)--(6.898,1.075);
\draw[gp path] (6.898,7.075)--(6.898,6.985);
\draw[gp path] (7.157,0.985)--(7.157,1.075);
\draw[gp path] (7.157,7.075)--(7.157,6.985);
\draw[gp path] (7.369,0.985)--(7.369,1.075);
\draw[gp path] (7.369,7.075)--(7.369,6.985);
\draw[gp path] (7.549,0.985)--(7.549,1.075);
\draw[gp path] (7.549,7.075)--(7.549,6.985);
\draw[gp path] (7.704,0.985)--(7.704,1.075);
\draw[gp path] (7.704,7.075)--(7.704,6.985);
\draw[gp path] (7.841,0.985)--(7.841,1.075);
\draw[gp path] (7.841,7.075)--(7.841,6.985);
\gpcolor{color=gp lt color axes}
\gpsetlinetype{gp lt axes}
\gpsetdashtype{gp dt axes}
\gpsetlinewidth{0.50}
\draw[gp path] (7.963,0.985)--(7.963,7.075);
\gpcolor{color=gp lt color border}
\gpsetlinetype{gp lt border}
\gpsetdashtype{gp dt solid}
\gpsetlinewidth{1.00}
\draw[gp path] (7.963,0.985)--(7.963,1.165);
\draw[gp path] (7.963,7.075)--(7.963,6.895);
\node[gp node center,font={\fontsize{8.0pt}{9.6pt}\selectfont}] at (7.963,0.677) {{      1e-06}};
\draw[gp path] (8.769,0.985)--(8.769,1.075);
\draw[gp path] (8.769,7.075)--(8.769,6.985);
\draw[gp path] (9.241,0.985)--(9.241,1.075);
\draw[gp path] (9.241,7.075)--(9.241,6.985);
\draw[gp path] (9.575,0.985)--(9.575,1.075);
\draw[gp path] (9.575,7.075)--(9.575,6.985);
\draw[gp path] (9.835,0.985)--(9.835,1.075);
\draw[gp path] (9.835,7.075)--(9.835,6.985);
\draw[gp path] (10.047,0.985)--(10.047,1.075);
\draw[gp path] (10.047,7.075)--(10.047,6.985);
\draw[gp path] (10.226,0.985)--(10.226,1.075);
\draw[gp path] (10.226,7.075)--(10.226,6.985);
\draw[gp path] (10.381,0.985)--(10.381,1.075);
\draw[gp path] (10.381,7.075)--(10.381,6.985);
\draw[gp path] (10.518,0.985)--(10.518,1.075);
\draw[gp path] (10.518,7.075)--(10.518,6.985);
\gpcolor{color=gp lt color axes}
\gpsetlinetype{gp lt axes}
\gpsetdashtype{gp dt axes}
\gpsetlinewidth{0.50}
\draw[gp path] (10.641,0.985)--(10.641,7.075);
\gpcolor{color=gp lt color border}
\gpsetlinetype{gp lt border}
\gpsetdashtype{gp dt solid}
\gpsetlinewidth{1.00}
\draw[gp path] (10.641,0.985)--(10.641,1.165);
\draw[gp path] (10.641,7.075)--(10.641,6.895);
\node[gp node center,font={\fontsize{8.0pt}{9.6pt}\selectfont}] at (10.641,0.677) {{      1e-05}};
\draw[gp path] (11.447,0.985)--(11.447,1.075);
\draw[gp path] (11.447,7.075)--(11.447,6.985);
\draw[gp path] (2.608,7.075)--(2.608,0.985)--(11.447,0.985)--(11.447,7.075)--cycle;
\node[gp node center,rotate=-270,font={\fontsize{10.0pt}{12.0pt}\selectfont}] at (1.196,4.030) {$\frac{|\mathcal{J}(\vE)-\mathcal{J}(\vE^*)|}{\mathcal{J}(\vE^*)}$};
\node[gp node center,font={\fontsize{8.0pt}{9.6pt}\selectfont}] at (7.027,0.215) {1/DoF};
\node[gp node center] at (7.027,7.537) {Convergence rate};
\node[gp node right,font={\fontsize{8.0pt}{9.6pt}\selectfont}] at (6.026,6.772) {uniform refinement};
\gpcolor{rgb color={0.580,0.000,0.827}}
\gpsetpointsize{4.00}
\gppoint{gp mark 1}{(10.869,6.603)}
\gppoint{gp mark 1}{(9.259,6.153)}
\gppoint{gp mark 1}{(7.648,5.721)}
\gppoint{gp mark 1}{(6.036,5.303)}
\gppoint{gp mark 1}{(4.424,4.869)}
\gppoint{gp mark 1}{(2.812,4.369)}
\gppoint{gp mark 1}{(6.557,6.772)}
\gpcolor{color=gp lt color border}
\node[gp node right,font={\fontsize{8.0pt}{9.6pt}\selectfont}] at (6.026,6.526) {local refinement};
\gpcolor{rgb color={0.000,0.620,0.451}}
\gppoint{gp mark 2}{(10.869,6.609)}
\gppoint{gp mark 2}{(10.823,6.480)}
\gppoint{gp mark 2}{(10.740,6.160)}
\gppoint{gp mark 2}{(10.605,5.777)}
\gppoint{gp mark 2}{(10.389,5.972)}
\gppoint{gp mark 2}{(10.107,5.589)}
\gppoint{gp mark 2}{(9.753,5.487)}
\gppoint{gp mark 2}{(9.338,5.065)}
\gppoint{gp mark 2}{(8.852,4.674)}
\gppoint{gp mark 2}{(8.338,4.252)}
\gppoint{gp mark 2}{(7.810,3.853)}
\gppoint{gp mark 2}{(7.271,3.522)}
\gppoint{gp mark 2}{(6.728,3.111)}
\gppoint{gp mark 2}{(6.188,2.788)}
\gppoint{gp mark 2}{(5.649,2.434)}
\gppoint{gp mark 2}{(5.114,1.694)}
\gppoint{gp mark 2}{(4.057,1.331)}
\gppoint{gp mark 2}{(6.557,6.526)}
\gpcolor{color=gp lt color border}
\node[gp node right,font={\fontsize{8.0pt}{9.6pt}\selectfont}] at (6.026,6.280) {uniform refinement fit};
\gpcolor{rgb color={0.337,0.706,0.914}}
\draw[gp path] (6.173,6.280)--(6.941,6.280);
\draw[gp path] (2.608,4.328)--(2.697,4.353)--(2.787,4.378)--(2.876,4.402)--(2.965,4.427)%
  --(3.054,4.451)--(3.144,4.476)--(3.233,4.500)--(3.322,4.525)--(3.412,4.550)--(3.501,4.574)%
  --(3.590,4.599)--(3.679,4.623)--(3.769,4.648)--(3.858,4.672)--(3.947,4.697)--(4.037,4.722)%
  --(4.126,4.746)--(4.215,4.771)--(4.304,4.795)--(4.394,4.820)--(4.483,4.845)--(4.572,4.869)%
  --(4.662,4.894)--(4.751,4.918)--(4.840,4.943)--(4.929,4.967)--(5.019,4.992)--(5.108,5.017)%
  --(5.197,5.041)--(5.286,5.066)--(5.376,5.090)--(5.465,5.115)--(5.554,5.140)--(5.644,5.164)%
  --(5.733,5.189)--(5.822,5.213)--(5.911,5.238)--(6.001,5.262)--(6.090,5.287)--(6.179,5.312)%
  --(6.269,5.336)--(6.358,5.361)--(6.447,5.385)--(6.536,5.410)--(6.626,5.434)--(6.715,5.459)%
  --(6.804,5.484)--(6.894,5.508)--(6.983,5.533)--(7.072,5.557)--(7.161,5.582)--(7.251,5.607)%
  --(7.340,5.631)--(7.429,5.656)--(7.519,5.680)--(7.608,5.705)--(7.697,5.729)--(7.786,5.754)%
  --(7.876,5.779)--(7.965,5.803)--(8.054,5.828)--(8.144,5.852)--(8.233,5.877)--(8.322,5.901)%
  --(8.411,5.926)--(8.501,5.951)--(8.590,5.975)--(8.679,6.000)--(8.769,6.024)--(8.858,6.049)%
  --(8.947,6.074)--(9.036,6.098)--(9.126,6.123)--(9.215,6.147)--(9.304,6.172)--(9.393,6.196)%
  --(9.483,6.221)--(9.572,6.246)--(9.661,6.270)--(9.751,6.295)--(9.840,6.319)--(9.929,6.344)%
  --(10.018,6.368)--(10.108,6.393)--(10.197,6.418)--(10.286,6.442)--(10.376,6.467)--(10.465,6.491)%
  --(10.554,6.516)--(10.643,6.541)--(10.733,6.565)--(10.822,6.590)--(10.911,6.614)--(11.001,6.639)%
  --(11.090,6.663)--(11.179,6.688)--(11.268,6.713)--(11.358,6.737)--(11.447,6.762);
\gpcolor{color=gp lt color border}
\node[gp node right,font={\fontsize{8.0pt}{9.6pt}\selectfont}] at (6.026,6.034) {local refinement fit};
\gpcolor{rgb color={0.902,0.624,0.000}}
\draw[gp path] (6.173,6.034)--(6.941,6.034);
\draw[gp path] (3.739,0.985)--(3.769,1.006)--(3.858,1.070)--(3.947,1.134)--(4.037,1.198)%
  --(4.126,1.262)--(4.215,1.326)--(4.304,1.390)--(4.394,1.454)--(4.483,1.518)--(4.572,1.583)%
  --(4.662,1.647)--(4.751,1.711)--(4.840,1.775)--(4.929,1.839)--(5.019,1.903)--(5.108,1.967)%
  --(5.197,2.031)--(5.286,2.095)--(5.376,2.159)--(5.465,2.223)--(5.554,2.288)--(5.644,2.352)%
  --(5.733,2.416)--(5.822,2.480)--(5.911,2.544)--(6.001,2.608)--(6.090,2.672)--(6.179,2.736)%
  --(6.269,2.800)--(6.358,2.864)--(6.447,2.928)--(6.536,2.993)--(6.626,3.057)--(6.715,3.121)%
  --(6.804,3.185)--(6.894,3.249)--(6.983,3.313)--(7.072,3.377)--(7.161,3.441)--(7.251,3.505)%
  --(7.340,3.569)--(7.429,3.633)--(7.519,3.698)--(7.608,3.762)--(7.697,3.826)--(7.786,3.890)%
  --(7.876,3.954)--(7.965,4.018)--(8.054,4.082)--(8.144,4.146)--(8.233,4.210)--(8.322,4.274)%
  --(8.411,4.338)--(8.501,4.403)--(8.590,4.467)--(8.679,4.531)--(8.769,4.595)--(8.858,4.659)%
  --(8.947,4.723)--(9.036,4.787)--(9.126,4.851)--(9.215,4.915)--(9.304,4.979)--(9.393,5.043)%
  --(9.483,5.108)--(9.572,5.172)--(9.661,5.236)--(9.751,5.300)--(9.840,5.364)--(9.929,5.428)%
  --(10.018,5.492)--(10.108,5.556)--(10.197,5.620)--(10.286,5.684)--(10.376,5.748)--(10.465,5.813)%
  --(10.554,5.877)--(10.643,5.941)--(10.733,6.005)--(10.822,6.069)--(10.911,6.133)--(11.001,6.197)%
  --(11.090,6.261)--(11.179,6.325)--(11.268,6.389)--(11.358,6.453)--(11.447,6.518);
\gpcolor{color=gp lt color border}
\draw[gp path] (2.608,7.075)--(2.608,0.985)--(11.447,0.985)--(11.447,7.075)--cycle;
\gpdefrectangularnode{gp plot 1}{\pgfpoint{2.608cm}{0.985cm}}{\pgfpoint{11.447cm}{7.075cm}}
\end{tikzpicture}

%% file: dual-dual.tex
\begin{tikzpicture}[gnuplot, scale=0.8]
\path (0.000,0.000) rectangle (13.000,4.000);
\gpcolor{color=gp lt color axes}
\gpsetlinetype{gp lt axes}
\gpsetdashtype{gp dt axes}
\gpsetlinewidth{0.50}
\draw[gp path] (1.962,0.031)--(11.805,0.031);
\gpcolor{color=gp lt color border}
\gpsetlinetype{gp lt border}
\gpsetdashtype{gp dt solid}
\gpsetlinewidth{1.00}
\draw[gp path] (1.962,0.031)--(2.142,0.031);
\draw[gp path] (11.805,0.031)--(11.625,0.031);
\node[gp node right] at (1.778,0.031) {$0.2$};
\gpcolor{color=gp lt color axes}
\gpsetlinetype{gp lt axes}
\gpsetdashtype{gp dt axes}
\gpsetlinewidth{0.50}
\draw[gp path] (1.962,0.687)--(11.805,0.687);
\gpcolor{color=gp lt color border}
\gpsetlinetype{gp lt border}
\gpsetdashtype{gp dt solid}
\gpsetlinewidth{1.00}
\draw[gp path] (1.962,0.687)--(2.142,0.687);
\draw[gp path] (11.805,0.687)--(11.625,0.687);
\node[gp node right] at (1.778,0.687) {$0.3$};
\gpcolor{color=gp lt color axes}
\gpsetlinetype{gp lt axes}
\gpsetdashtype{gp dt axes}
\gpsetlinewidth{0.50}
\draw[gp path] (1.962,1.343)--(11.805,1.343);
\gpcolor{color=gp lt color border}
\gpsetlinetype{gp lt border}
\gpsetdashtype{gp dt solid}
\gpsetlinewidth{1.00}
\draw[gp path] (1.962,1.343)--(2.142,1.343);
\draw[gp path] (11.805,1.343)--(11.625,1.343);
\node[gp node right] at (1.778,1.343) {$0.4$};
\gpcolor{color=gp lt color axes}
\gpsetlinetype{gp lt axes}
\gpsetdashtype{gp dt axes}
\gpsetlinewidth{0.50}
\draw[gp path] (1.962,1.999)--(11.805,1.999);
\gpcolor{color=gp lt color border}
\gpsetlinetype{gp lt border}
\gpsetdashtype{gp dt solid}
\gpsetlinewidth{1.00}
\draw[gp path] (1.962,1.999)--(2.142,1.999);
\draw[gp path] (11.805,1.999)--(11.625,1.999);
\node[gp node right] at (1.778,1.999) {$0.5$};
\gpcolor{color=gp lt color axes}
\gpsetlinetype{gp lt axes}
\gpsetdashtype{gp dt axes}
\gpsetlinewidth{0.50}
\draw[gp path] (1.962,2.656)--(11.805,2.656);
\gpcolor{color=gp lt color border}
\gpsetlinetype{gp lt border}
\gpsetdashtype{gp dt solid}
\gpsetlinewidth{1.00}
\draw[gp path] (1.962,2.656)--(2.142,2.656);
\draw[gp path] (11.805,2.656)--(11.625,2.656);
\node[gp node right] at (1.778,2.656) {$0.6$};
\gpcolor{color=gp lt color axes}
\gpsetlinetype{gp lt axes}
\gpsetdashtype{gp dt axes}
\gpsetlinewidth{0.50}
\draw[gp path] (1.962,3.312)--(11.805,3.312);
\gpcolor{color=gp lt color border}
\gpsetlinetype{gp lt border}
\gpsetdashtype{gp dt solid}
\gpsetlinewidth{1.00}
\draw[gp path] (1.962,3.312)--(2.142,3.312);
\draw[gp path] (11.805,3.312)--(11.625,3.312);
\node[gp node right] at (1.778,3.312) {$0.7$};
\gpcolor{color=gp lt color axes}
\gpsetlinetype{gp lt axes}
\gpsetdashtype{gp dt axes}
\gpsetlinewidth{0.50}
\draw[gp path] (1.962,3.968)--(11.805,3.968);
\gpcolor{color=gp lt color border}
\gpsetlinetype{gp lt border}
\gpsetdashtype{gp dt solid}
\gpsetlinewidth{1.00}
\draw[gp path] (1.962,3.968)--(2.142,3.968);
\draw[gp path] (11.805,3.968)--(11.625,3.968);
\node[gp node right] at (1.778,3.968) {$0.8$};
\gpcolor{color=gp lt color axes}
\gpsetlinetype{gp lt axes}
\gpsetdashtype{gp dt axes}
\gpsetlinewidth{0.50}
\draw[gp path] (2.719,0.031)--(2.719,3.968);
\gpcolor{color=gp lt color border}
\gpsetlinetype{gp lt border}
\gpsetdashtype{gp dt solid}
\gpsetlinewidth{1.00}
\draw[gp path] (2.719,0.031)--(2.719,0.211);
\draw[gp path] (2.719,3.968)--(2.719,3.788);
\node[gp node center] at (2.719,-0.277) {$0.02$};
\gpcolor{color=gp lt color axes}
\gpsetlinetype{gp lt axes}
\gpsetdashtype{gp dt axes}
\gpsetlinewidth{0.50}
\draw[gp path] (3.729,0.031)--(3.729,3.968);
\gpcolor{color=gp lt color border}
\gpsetlinetype{gp lt border}
\gpsetdashtype{gp dt solid}
\gpsetlinewidth{1.00}
\draw[gp path] (3.729,0.031)--(3.729,0.211);
\draw[gp path] (3.729,3.968)--(3.729,3.788);
\node[gp node center] at (3.729,-0.277) {$0.04$};
\gpcolor{color=gp lt color axes}
\gpsetlinetype{gp lt axes}
\gpsetdashtype{gp dt axes}
\gpsetlinewidth{0.50}
\draw[gp path] (4.738,0.031)--(4.738,3.968);
\gpcolor{color=gp lt color border}
\gpsetlinetype{gp lt border}
\gpsetdashtype{gp dt solid}
\gpsetlinewidth{1.00}
\draw[gp path] (4.738,0.031)--(4.738,0.211);
\draw[gp path] (4.738,3.968)--(4.738,3.788);
\node[gp node center] at (4.738,-0.277) {$0.06$};
\gpcolor{color=gp lt color axes}
\gpsetlinetype{gp lt axes}
\gpsetdashtype{gp dt axes}
\gpsetlinewidth{0.50}
\draw[gp path] (5.748,0.031)--(5.748,3.968);
\gpcolor{color=gp lt color border}
\gpsetlinetype{gp lt border}
\gpsetdashtype{gp dt solid}
\gpsetlinewidth{1.00}
\draw[gp path] (5.748,0.031)--(5.748,0.211);
\draw[gp path] (5.748,3.968)--(5.748,3.788);
\node[gp node center] at (5.748,-0.277) {$0.08$};
\gpcolor{color=gp lt color axes}
\gpsetlinetype{gp lt axes}
\gpsetdashtype{gp dt axes}
\gpsetlinewidth{0.50}
\draw[gp path] (6.757,0.031)--(6.757,3.968);
\gpcolor{color=gp lt color border}
\gpsetlinetype{gp lt border}
\gpsetdashtype{gp dt solid}
\gpsetlinewidth{1.00}
\draw[gp path] (6.757,0.031)--(6.757,0.211);
\draw[gp path] (6.757,3.968)--(6.757,3.788);
\node[gp node center] at (6.757,-0.277) {$0.1$};
\gpcolor{color=gp lt color axes}
\gpsetlinetype{gp lt axes}
\gpsetdashtype{gp dt axes}
\gpsetlinewidth{0.50}
\draw[gp path] (7.767,0.031)--(7.767,3.480);
\draw[gp path] (7.767,3.788)--(7.767,3.968);
\gpcolor{color=gp lt color border}
\gpsetlinetype{gp lt border}
\gpsetdashtype{gp dt solid}
\gpsetlinewidth{1.00}
\draw[gp path] (7.767,0.031)--(7.767,0.211);
\draw[gp path] (7.767,3.968)--(7.767,3.788);
\node[gp node center] at (7.767,-0.277) {$0.12$};
\gpcolor{color=gp lt color axes}
\gpsetlinetype{gp lt axes}
\gpsetdashtype{gp dt axes}
\gpsetlinewidth{0.50}
\draw[gp path] (8.776,0.031)--(8.776,3.480);
\draw[gp path] (8.776,3.788)--(8.776,3.968);
\gpcolor{color=gp lt color border}
\gpsetlinetype{gp lt border}
\gpsetdashtype{gp dt solid}
\gpsetlinewidth{1.00}
\draw[gp path] (8.776,0.031)--(8.776,0.211);
\draw[gp path] (8.776,3.968)--(8.776,3.788);
\node[gp node center] at (8.776,-0.277) {$0.14$};
\gpcolor{color=gp lt color axes}
\gpsetlinetype{gp lt axes}
\gpsetdashtype{gp dt axes}
\gpsetlinewidth{0.50}
\draw[gp path] (9.786,0.031)--(9.786,3.480);
\draw[gp path] (9.786,3.788)--(9.786,3.968);
\gpcolor{color=gp lt color border}
\gpsetlinetype{gp lt border}
\gpsetdashtype{gp dt solid}
\gpsetlinewidth{1.00}
\draw[gp path] (9.786,0.031)--(9.786,0.211);
\draw[gp path] (9.786,3.968)--(9.786,3.788);
\node[gp node center] at (9.786,-0.277) {$0.16$};
\gpcolor{color=gp lt color axes}
\gpsetlinetype{gp lt axes}
\gpsetdashtype{gp dt axes}
\gpsetlinewidth{0.50}
\draw[gp path] (10.795,0.031)--(10.795,3.480);
\draw[gp path] (10.795,3.788)--(10.795,3.968);
\gpcolor{color=gp lt color border}
\gpsetlinetype{gp lt border}
\gpsetdashtype{gp dt solid}
\gpsetlinewidth{1.00}
\draw[gp path] (10.795,0.031)--(10.795,0.211);
\draw[gp path] (10.795,3.968)--(10.795,3.788);
\node[gp node center] at (10.795,-0.277) {$0.18$};
\gpcolor{color=gp lt color axes}
\gpsetlinetype{gp lt axes}
\gpsetdashtype{gp dt axes}
\gpsetlinewidth{0.50}
\draw[gp path] (11.805,0.031)--(11.805,3.968);
\gpcolor{color=gp lt color border}
\gpsetlinetype{gp lt border}
\gpsetdashtype{gp dt solid}
\gpsetlinewidth{1.00}
\draw[gp path] (11.805,0.031)--(11.805,0.211);
\draw[gp path] (11.805,3.968)--(11.805,3.788);
\node[gp node center] at (11.805,-0.277) {$0.2$};
\draw[gp path] (1.962,3.968)--(1.962,0.031)--(11.805,0.031)--(11.805,3.968)--cycle;

\node[gp node left] at (4.385,0.803) {$d_{\,\text{opt}}$};
\gpsetlinewidth{2.00}
\gpsetdashtype{gp dt 3}
\draw[gp path](4.357,3.968)--(4.357,0.031);
\node[gp node center,rotate=-270] at (0.488,1.999) {\small Ratio};
\node[gp node center] at (6.883,-0.838) {\small Interlayer height, $d$};
\node[gp node right] at (10.337,1.134) {\small double layer system};
\node[gp node right] at (10.337,0.634) {\small single layer system};
\gpcolor{rgb color={0.000,0.620,0.451}}
\gpsetlinewidth{2.00}
\gpsetdashtype{gp dt 2}
\draw[gp path](1.962,2.196)--(11.805,2.196);
\draw[gp path] (10.521,0.634)--(11.437,0.634);
\gpcolor{rgb color={0.580,0.000,0.827}}
\gpsetdashtype{gp dt solid}
\draw[gp path] (10.521,1.134)--(11.437,1.134);
\draw[gp path] (1.962,1.962)--(1.992,2.084)--(2.028,2.190)--(2.058,2.284)--(2.093,2.373)%
  --(2.129,2.438)--(2.159,2.508)--(2.189,2.562)--(2.224,2.618)--(2.260,2.669)--(2.290,2.713)%
  --(2.320,2.757)--(2.356,2.797)--(2.391,2.834)--(2.421,2.868)--(2.452,2.899)--(2.487,2.930)%
  --(2.517,2.958)--(2.553,2.985)--(2.588,3.009)--(2.618,3.033)--(2.648,3.055)--(2.684,3.077)%
  --(2.714,3.096)--(2.749,3.114)--(2.785,3.132)--(2.815,3.149)--(2.850,3.165)--(2.881,3.181)%
  --(2.911,3.195)--(2.946,3.208)--(2.977,3.221)--(3.012,3.233)--(3.047,3.245)--(3.078,3.256)%
  --(3.113,3.267)--(3.143,3.277)--(3.173,3.286)--(3.209,3.295)--(3.239,3.304)--(3.274,3.312)%
  --(3.305,3.320)--(3.340,3.328)--(3.375,3.335)--(3.406,3.342)--(3.441,3.348)--(3.471,3.354)%
  --(3.502,3.360)--(3.537,3.365)--(3.572,3.370)--(3.603,3.375)--(3.638,3.380)--(3.668,3.384)%
  --(3.698,3.387)--(3.734,3.391)--(3.769,3.394)--(3.799,3.397)--(3.830,3.400)--(3.865,3.403)%
  --(3.895,3.406)--(3.931,3.409)--(3.966,3.411)--(3.996,3.413)--(4.027,3.415)--(4.062,3.417)%
  --(4.097,3.418)--(4.127,3.419)--(4.163,3.421)--(4.193,3.422)--(4.223,3.422)--(4.259,3.423)%
  --(4.294,3.423)--(4.324,3.424)--(4.355,3.424)--(4.390,3.424)--(4.420,3.424)--(4.456,3.423)%
  --(4.491,3.423)--(4.521,3.422)--(4.551,3.421)--(4.587,3.420)--(4.617,3.419)--(4.652,3.418)%
  --(4.688,3.417)--(4.718,3.415)--(4.748,3.414)--(4.784,3.412)--(4.819,3.410)--(4.849,3.408)%
  --(4.885,3.406)--(4.915,3.403)--(4.945,3.401)--(4.981,3.398)--(5.011,3.395)--(5.046,3.393)%
  --(5.081,3.390)--(5.112,3.386)--(5.142,3.383)--(5.177,3.380)--(5.208,3.376)--(5.243,3.373)%
  --(5.309,3.365)--(5.374,3.357)--(5.440,3.349)--(5.505,3.339)--(5.571,3.329)--(5.637,3.319)%
  --(5.702,3.308)--(5.768,3.297)--(5.834,3.286)--(5.899,3.274)--(5.965,3.262)--(6.030,3.250)%
  --(6.096,3.237)--(6.162,3.224)--(6.227,3.211)--(6.293,3.197)--(6.359,3.183)--(6.424,3.169)%
  --(6.490,3.155)--(6.555,3.141)--(6.621,3.126)--(6.687,3.111)--(6.752,3.096)--(6.818,3.081)%
  --(6.884,3.066)--(6.949,3.051)--(7.015,3.035)--(7.080,3.020)--(7.146,3.005)--(7.212,2.989)%
  --(7.277,2.974)--(7.343,2.958)--(7.408,2.942)--(7.474,2.928)--(7.540,2.912)--(7.605,2.897)%
  --(7.671,2.882)--(7.737,2.867)--(7.802,2.852)--(7.868,2.837)--(7.933,2.822)--(7.999,2.807)%
  --(8.065,2.792)--(8.130,2.778)--(8.196,2.764)--(8.262,2.749)--(8.327,2.736)--(8.393,2.721)%
  --(8.458,2.707)--(8.524,2.694)--(8.590,2.680)--(8.655,2.667)--(8.721,2.654)--(8.786,2.641)%
  --(8.852,2.628)--(8.918,2.616)--(8.983,2.603)--(9.049,2.591)--(9.115,2.579)--(9.180,2.567)%
  --(9.246,2.555)--(9.311,2.544)--(9.377,2.532)--(9.443,2.521)--(9.508,2.510)--(9.574,2.499)%
  --(9.640,2.489)--(9.705,2.478)--(9.771,2.468)--(9.836,2.458)--(9.902,2.448)--(9.968,2.438)%
  --(10.033,2.428)--(10.099,2.419)--(10.165,2.410)--(10.230,2.401)--(10.296,2.392)--(10.361,2.383)%
  --(10.427,2.374)--(10.493,2.366)--(10.558,2.358)--(10.624,2.349)--(10.689,2.341)--(10.755,2.334)%
  --(10.821,2.326)--(10.886,2.318)--(10.952,2.311)--(11.018,2.304)--(11.083,2.297)--(11.149,2.290)%
  --(11.214,2.283)--(11.280,2.276)--(11.346,2.270)--(11.411,2.263)--(11.477,2.257)--(11.543,2.250)%
  --(11.608,2.244)--(11.674,2.238)--(11.739,2.233)--(11.805,2.227);
\gpcolor{color=gp lt color border}
\gpsetlinewidth{1.00}
\draw[gp path] (1.962,3.968)--(1.962,0.031)--(11.805,0.031)--(11.805,3.968)--cycle;
\gpdefrectangularnode{gp plot 1}{\pgfpoint{1.962cm}{0.031cm}}{\pgfpoint{11.805cm}{3.968cm}}
\end{tikzpicture}